\documentclass[11pt]{article}

\usepackage[preprint]{acl}

\usepackage{times}
\usepackage{latexsym}

\usepackage[T1]{fontenc}

\usepackage[utf8]{inputenc}

\usepackage{microtype}

\usepackage{inconsolata}

\usepackage{graphicx}

\usepackage{amsmath}
\usepackage{amsfonts}
\usepackage{booktabs}
\usepackage{algorithm}
\usepackage{algpseudocode}
\usepackage{caption}
\usepackage{tcolorbox}
\usepackage{caption}
\usepackage{subcaption}     
\usepackage{multirow}
\usepackage{array} 

\usepackage{algorithm}
\usepackage{algpseudocode}
\title{Zipage: Maintain High Request Concurrency for LLM Reasoning through Compressed PagedAttention}

\author{
Mengqi Liao$^{1,2}$ , Lu Wang$^{2}$, Chaoyun Zhang$^{2}$, Bo Qiao$^{2}$, \\
\textbf{Si Qin}$^{2}$, \textbf{Qingwei Lin}$^{2}$, \textbf{Saravan Rajmohan}$^{2}$, \textbf{Dongmei Zhang}$^{2}$ \& \textbf{Huaiyu Wan}$^{1,3,*}$
\\
$^1$School of Computer Science and Technology, Beijing Jiaotong University \\
$^2$Microsoft  \\
$^3$Beijing Key Laboratory of Traffic Data Mining and Embodied Intelligence \\
\\
\texttt{mqliao@bjtu.edu.cn, wlu@microsoft.com, hywan@bjtu.edu.cn} 
}

\begin{document}
\maketitle
\begin{abstract}

With reasoning becoming the generative paradigm for large language models (LLMs),  the memory bottleneck caused by KV cache during the decoding phase has become a critical factor limiting high-concurrency service. Although existing KV cache eviction methods address the memory issue, most of them are impractical for industrial-grade applications. This paper introduces Compressed PagedAttention, a method that combines token-wise KV cache eviction with PagedAttention. We propose a comprehensive scheduling strategy and support prefix caching and asynchronous compression for Compressed PagedAttention. Based on this, we have developed a high-concurrency LLM inference engine, Zipage. On large-scale mathematical reasoning tasks, Zipage achieves around 95\% of the performance of Full KV inference engines while delivering over 2.1$\times$ speedup.

\end{abstract}

\section{Introduction}

With the advancement of large language models (LLMs), reasoning LLMs have garnered increasing attention from the community \citep{ke2025survey,li2025system}. These models typically perform extensive reasoning before generating answers and have shown remarkable progress in complex domains like code and mathematics. However, as sequence length grows, the memory required for storing the KV cache increases significantly. The core bottleneck of LLM service systems has shifted from computation to having sufficient memory to sustain high-concurrency execution in long sequence scenarios.

Existing KV cache eviction methods can reduce memory usage at the algorithmic level but face fundamental mismatches at the system level. While some methods \citep{morghkv,rkv,gkv} achieve constant memory usage during decoding, they lack support for advanced techniques like continuous batching and prefix caching, essential features in modern inference engines such as vLLM\footnote{https://github.com/vllm-project/vllm} and SGLang\footnote{https://github.com/sgl-project/sglang}. Consequently, their actual throughput is often lower than engines using a full KV cache. Other methods integrate KV cache eviction into inference engines but rely on \textbf{coarse-grained page-wise} eviction, risking the loss of critical information and degrading performance \cite{raas, pagedeviction}. KV-Compress \citep{rehg2024kv}, though employing token-wise eviction, only supports input compression and \textbf{disrupts the prefix cache}, significantly increasing prefilling costs.

In this paper, we propose Compressed PagedAttention, a KV cache management approach that combines PagedAttention \citep{kwon2023efficient}  with \textbf{flexible token-wise KV cache eviction  across layers and attention heads}.  We implemented a high-concurrency inference engine, Zipage\footnote{
https://github.com/microsoft/Zipage
}, based on Compressed PagedAttention, and developed \textbf{efficient GPU kernels} to optimize operations during the compression process.  Zipage employs a comprehensive request scheduling strategy designed for Compressed PagedAttention and is \textbf{compatible with prefix caching}, achieving significant throughput improvements in scenarios where many requests share the same prefix. It also implements \textbf{asynchronous compression and decoding} to further enhance throughput. 

\begin{figure*}[!ht]
    \centering
    \includegraphics[width=0.85\linewidth, trim=30 100 20 70, clip]{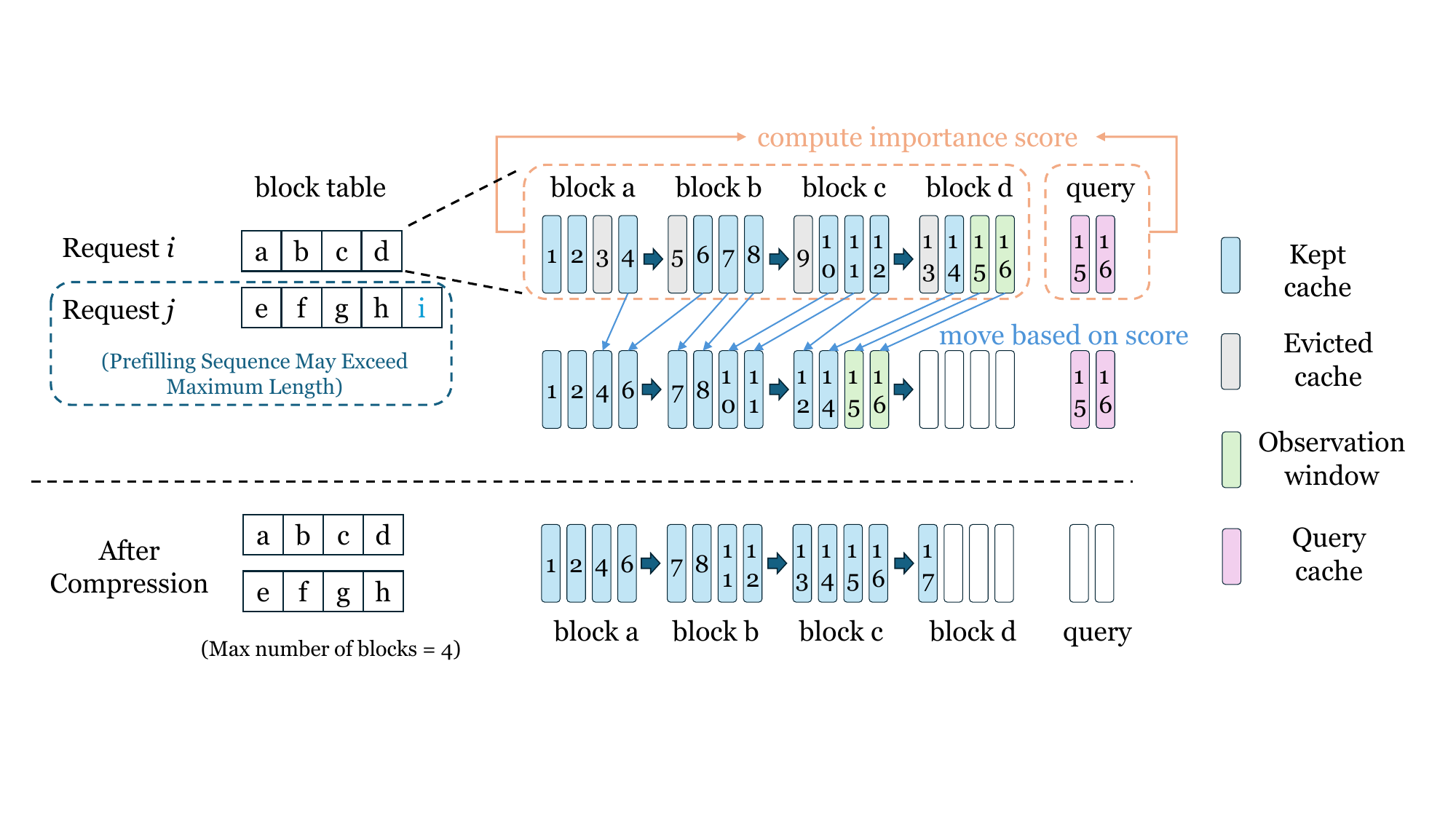}
    \caption{ Illustration of Compressed PagedAttention. Here, $N_{\max}=4, b=4, w=2$. The figure depicts two requests requiring compression. After compression, the kept KV cache entries are moved to the first three blocks, while the fourth block is reserved for subsequent decoding. The remaining blocks are released.}
    \label{fig:compressed_pagedattention}
\end{figure*}

We evaluated models of various architectures and sizes on reasoning tasks, such as coding and mathematics. Zipage achieved significant throughput gains while maintaining performance close to a Full KV cache engine. Specifically, in mathematical reasoning tasks, Zipage achieves over a 2.1\(\times\) speedup while retaining approximately 95\% of the performance of a Full KV cache engine.

\section{Related Work}

Some methods reduce computational complexity and memory usage by evicting KV cache entries. For example, SnapKV \cite{li2024snapkv} calculates attention scores to decide which token' KV cache to retain or evict, while PyramidInfer \cite{yang2024pyramidinfer}, PyramidKV \cite{cai2024pyramidkv}, and Ada-KV \cite{feng2024ada} adjust budgets across heads or layers to improve performance. However, these methods focus on compressing input KV cache, while output KV cache dominates memory in reasoning, limiting concurrency.

Methods such as MorphKV \citep{morghkv}, R-KV \citep{rkv}, and  G-KV \citep{gkv} perform KV cache eviction during the decoding process, ensuring that the KV cache for each request remains constant. Although these methods significantly improve concurrency, they are not integrated with inference engines and thus cannot be practically applied. 

\section{Background}

PagedAttention \citep{kwon2023efficient} is an efficient method for managing the KV cache of Transformer \citep{vaswani2017attention} based LLMs. The LLM serving engine, vLLM, is built on PagedAttention. 

\textbf{Pre-allocated memory.} vLLM pre-allocates GPU memory for the KV cache, denoted as $\mathbf{K}, \mathbf{V} \in \mathbb{R}^{L  \times N_{\text{total}} \times b \times h_{\text{kv}}\times d}$. Here, $L$ represents the number of layers of  LLMs, $N_{\text{total}}$ is the total number of blocks, $b$ corresponds to the block size, $h_{\text{kv}}$ denotes the number of attention heads, and $d$ is the attention dimension.

\textbf{KV cache management.} vLLM partitions the sequence of each request according to the block size. The KV cache of each block is then written into free blocks in $\mathbf{K}$ and $\mathbf{V}$. vLLM  maintains a block table to record the blocks occupied by each request and their corresponding order. 

\textbf{Request scheduling.} In vLLM, requests are categorized into two states: waiting and running, and these states are managed using two separate queues. At each decoding step, if sufficient blocks are available for prefilling, vLLM transfers requests from the front of the waiting queue to the running queue and performs prefilling. Otherwise, the requests in the running queue decode the next token. When the last block of a request is filled and new blocks cannot be allocated, the running requests preempt the blocks of the most recently added requests in the running queue. The preempted requests are then moved back to the front of the waiting queue. This scheduling process adheres to a \textbf{first-come, first-served (FCFS)} principle.

\section{Method}

In this work, we propose a novel KV cache management method, Compressed PagedAttention, and develop an LLM serving engine, Zipage.

\subsection{Compressed PagedAttention}

To address the concurrency challenges, we introduce Compressed PagedAttention, a KV cache management method based on PagedAttention and KV cache eviction strategies.  Compressed PagedAttention builds upon PagedAttention by introducing the following key features:
\begin{itemize}
    \item The number of blocks occupied by each request is capped at $N_{\max}$, except during the prefilling phase, where the prefilling length may temporarily exceed this limit.
    \item After each decoding step, if a request occupies $N$ blocks and satisfies $N \ge N_{\max}$ with the last block fully occupied, a compression operation is triggered to evict less important KV cache entries and relocate the retained ones to the first $N_{\max} - 1$ blocks. The $N_{\max}$-th block is reserved for subsequent decoding, while the remaining blocks are released.
\end{itemize}
Compressed PagedAttention ensures that the memory usage of each request remains within a fixed maximum limit throughout the decoding process, thereby maintaining  high concurrency. Figure \ref{fig:compressed_pagedattention} illustrates the KV cache management mechanism in Compressed PagedAttention.

\subsection{The Compression Process Pipeline}
In this section, we will further elaborate on the compression process. First, following SnapKV \citep{li2024snapkv} and MorphKV \citep{morghkv},  we take the query states of the last $w$ tokens in the final block of each request as  \textit{observation window}. To accommodate these query states, we pre-allocate memory as $\mathbf Q \in \mathbb{R}^{L \times M \times w \times h_{q} \times d}$, where $M$ represents the maximum concurrency, $h_{q}$ is the number of attention heads.
The maximum concurrency $M$ is subject to the following constraints:  
\begin{equation}  
\left\{
\begin{aligned}
    &m_{\text{kv}} \times N_{\text{total}} + M \times m_{\text{q}} \leq m_{\text{available}}, \\
    &M \leq \frac{N_{\text{total}}}{N_{\max}}, \\
    &M > 0, \quad N_{\text{total}} > 0,
\end{aligned}
\right.
\label{eq:max_c}
\end{equation}  
where $m_{\text{available}}$ denotes the total available memory, $m_{\text{kv}}$ represents the memory required for the KV cache of a block, and $m_{\text{q}}$ is the memory required to cache the query states of a request. This is a linear programming problem, and the maximum value of $M$ is achieved when  $M = \lfloor \frac{m_{\text{available}}}{m_{\text{kv}}\times N_{\max}+m_{\text{q}}}\rfloor$, at which point $N_{\text{total}} = \lfloor\frac{m_{\text{available}}}{m_{\text{kv}}+{m_{\text{q}}}/{N_{\max}}} \rfloor$.

When compression is triggered, a scoring function $\phi(\mathbf Q,\mathbf K, \mathcal I)$ is employed to assign a score for the KV cache entries of requests that require compression. Here, $\mathcal{I}$ represents additional information, such as the block tables and query slots indexes for these requests. In its basic form , this scoring function involves computing attention scores between the query states in $\mathbf{Q}$ and the key states in $\mathbf{K}$. Furthermore, R-KV \citep{rkv} introduces a redundancy score to evaluate the redundancy of the KV cache, while G-KV \citep{gkv} incorporates a global score to aggregate historical attention scores, providing a better assessment of long-term importance. We integrate these methods into our framework and \textbf{implement kernel-level optimizations specifically tailored for the paged KV cache.} Detailed algorithm and experimental results can be found in Appendices \ref{sct:attention_score}, \ref{sct:global_score} and \ref{sct:similarity_score}. 

After obtaining the final scores, we assign a score of $+\infty$ to the entries within the observation window to ensure they are always retained. Subsequently, the top-$k$ KV cache entries with the highest scores are retained, where $k = (N_{\max} - 1) \times b$, referred to as the KV cache budget. Compression is then performed by reorganizing the retained KV cache entries such that their placement in $\mathbf{K}$ and $\mathbf{V}$ becomes compact and contiguous in a page. The full compression algorithm is described in Appendix~\ref{sct:compression}.

Additionally, although the raw redundancy score from R-KV significantly improves the performance, its computational complexity is $\mathcal{O}(N^2 \times b^2)$, becoming the primary bottleneck in the compression process. To address this issue, we propose a novel \textbf{lightning redundancy score} with a reduced computational complexity of $\mathcal{O}(N \times b^2)$, which not only significantly accelerates the compression but also achieves better performance than the raw redundancy score. Detailed descriptions and experiments are provided in Appendix \ref{sct:lightning_similarity}.

\subsection{Hybrid Scheduling}
\label{sct:hybrid}
To implement a LLM inference engine, a scheduling strategy tailored to Compressed PagedAttention is also crucial.
For Compressed PagedAttention, each request requires the allocation of  query slots for compression. The number of requests that can be allocated query slots is constrained by the maximum concurrency $M$. \textbf{The simplest scheduling strategy is to restrict the concurrency to no more than $M$.} Since some requests may occupy more than $N_{\max}$ blocks, it is possible for requests with fewer than $N_{\max}$ blocks to become blocked when attempting to allocate new blocks, even if the concurrency does not exceed $M$. For requests occupying more than $N_{\max}$ blocks, no additional blocks need to be allocated, and the extra blocks are released after the first compression. At this point, the blocked requests can resume decoding. This scheduling strategy, therefore, enables scheduling \textbf{without preemption}, and we refer to it as \textbf{constrained scheduling}.

Although constrained scheduling is simple and avoids preemption, this strategy may lead to underutilization of KV cache blocks. When there are a large number of requests with short inputs, the number of blocks occupied by these requests is less than $N_{\max}$, resulting in significant block idleness due to the concurrency limit. This underutilization becomes more frequent in scenarios where only brief responses are required. 

\begin{figure}[t]
    \centering
    \includegraphics[width=0.85\linewidth, trim=170 100 150 80, clip]{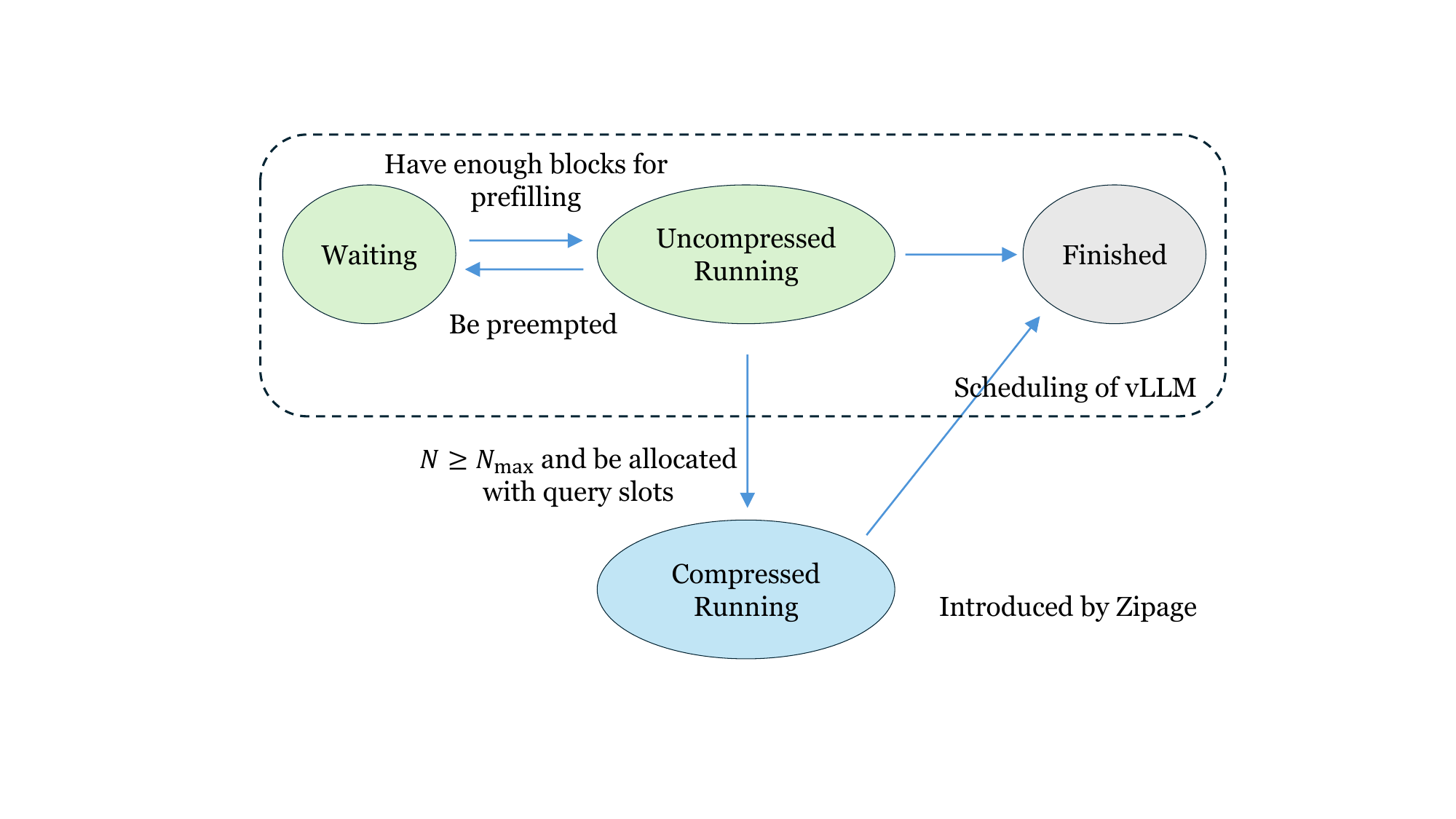}
    \caption{State transition diagram of requests under hybrid scheduling.}
    \label{fig:hybrid_state}
\end{figure}

To fully utilize these blocks and enhance concurrency, we propose a \textbf{hybrid scheduling} strategy. Specifically, the rules of hybrid scheduling can be summarized as follows:  
\begin{itemize}  
    \item Only the first $M$ requests in the running queue are eligible for query slots allocation.  
    \item Requests occupying fewer than $N_{\max}$ blocks or with fewer than $b-w$ tokens in the last block can decoding without being assigned query slots. Such requests from the waiting queue can be moved to the running queue for prefilling when sufficient blocks are available, even without query slot allocation. However, requests in the running queue without query slots will be blocked once they no longer meet these conditions. 
    \item When query slots are released, they are prioritally allocated to the foremost requests in the running queue that lack assigned query slots. 
    \item If a request in the running queue attempts to allocate a new block but no free blocks are available, preemption is triggered. Priority is given to offloading the last request without assigned query slots. Once all such requests are offloaded, the system \textbf{reverts to constrained scheduling}.
\end{itemize}

Under the hybrid scheduling strategy, the maximum concurrency is no longer constrained by $M$. The states of requests are illustrated in Figure \ref{fig:hybrid_state}. Requests that have already undergone compression can continue to run without preemption until completion, while uncompressed requests may be subject to preemption. Although the preempted requests discussed in this section are limited to those without assigned query slots, this rule will change in \S\ref{sct:share_prefix}, where all uncompressed requests, including those with assigned query slots, may be offloaded.

\subsection{Shared Prefix Cache for Compressed PagedAttention}
\label{sct:share_prefix}
Shared prefix cache is a key technique in inference engines. When multiple requests have the same prefix, the KV cache of the prefix can be shared across these requests. This approach reduces both memory usage and the computational overhead of prefilling.

\begin{figure}[t]
    \centering
    \includegraphics[width=0.9\linewidth, trim=170 120 300 50, clip]{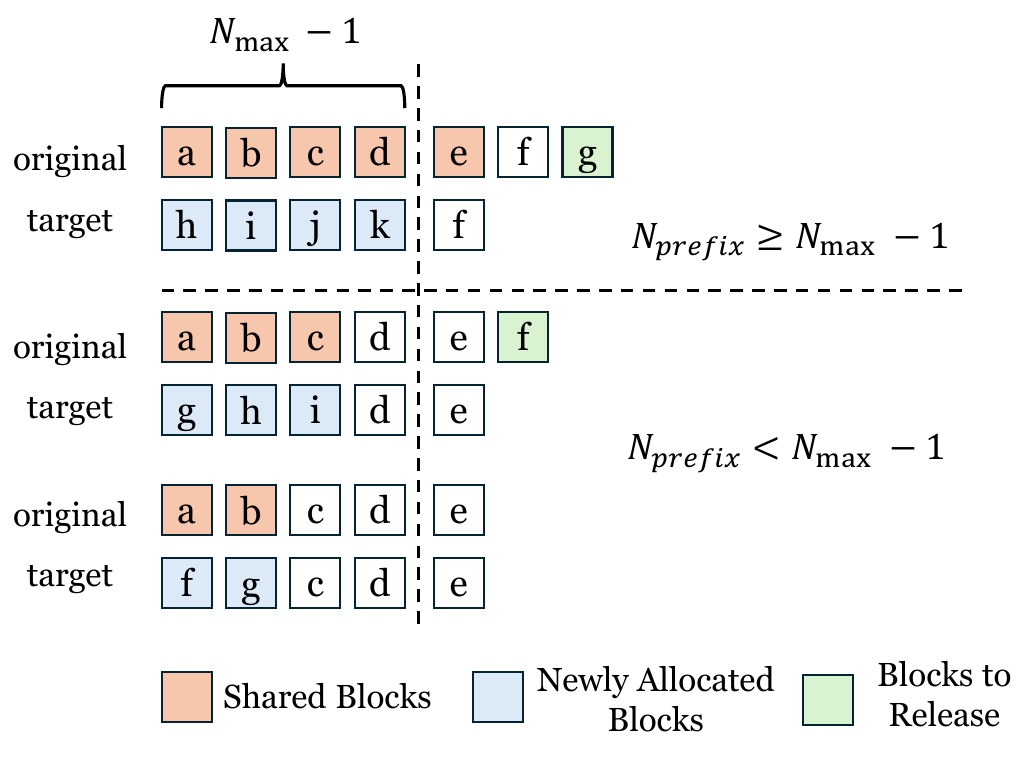}
    \caption{ Illustration of block allocation and release strategies for prefix cache.}
    \label{fig:prefix}
\end{figure}

In Compressed PagedAttention, the compression process will disrupt the shared prefix structure and different requests may retain different subsets of KV cache entries, making sharing infeasible. To resolve this, we modify the compression strategy: instead of rearranging KV cache entries within allocated blocks, compression is redirected to a set of target blocks, determined as follows:
\begin{itemize}
    \item Prefix caching is shared across requests at the block level. Each block tracks the number of requests referencing it. Blocks with a reference count greater than 1 are considered shared.
    \item If the number of shared  blocks for a request, denoted as \( N_{\text{prefix}} \), is greater than or equal to \( N_{\max} - 1 \), we allocate \( N_{\max} - 1 \) new blocks as taget blocks. The KV cache of the request is then compressed into these newly allocated blocks.
    \item If \( N_{\text{prefix}} < N_{\max} - 1 \), we allocate \( N_{\text{prefix}} \) new blocks and reuses \( N_{\max} - 1 - N_{\text{prefix}} \) blocks already allocated to the request. These combined blocks are used as the target blocks for compression.
\end{itemize}

With this adjustment, shared prefixes are preserved after compression. Figure~\ref{fig:prefix} illustrates examples of block tables before and after compression. As the compression is completed, the reference count for each shared block is decremented by 1. If the reference count drops to 1, the block is no longer considered a shared block.

Finally, as discussed in \S\ref{sct:hybrid}, under constrained scheduling, requests that attempt to allocate new blocks without availability would simply be blocked without releasing any blocks. However, with shared prefixes, new blocks may need to be allocated before compression, meaning requests occupying more than $N_{\max}$ blocks could also face blocking, \textbf{potentially leading to deadlocks}. To resolve this, preemption must be applied when prefix sharing is enabled. In such cases, the last \textbf{uncompressed request} will be preempted. Although this may occasionally deviate from the first-come, first-served principle, it still prevents prolonged request starvation.

\subsection{Asynchronous Decoding and Compression}

\begin{figure}[h]
    \centering
    \includegraphics[width=0.95\linewidth, trim=0 0 0 0, clip]{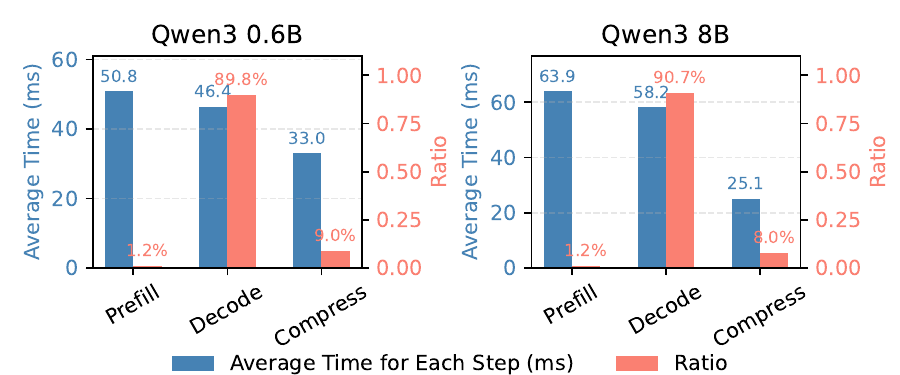}
    \caption{Average time per step and ratio during inference with Qwen3 0.6B and Qwen3 8B on AMC 23 under non-asynchronous compression settings.}
    \label{fig:three_stage}
\end{figure}

The previous section introduced the core concepts of Zipage. Here, we evaluate Zipage's performance on reasoning tasks. We measured the average time per step and its proportion of the total time spent on prefilling, decoding, and compression. As illustrated in Figure \ref{fig:three_stage}, decoding dominates the overall time consumption in reasoning tasks, while compression accounts for about 10\% of the total time. Additionally, the time required for each compression step is approximately $40\%-70\%$ of that for a decoding step.

We also observe that requests requiring compression constitute less than 1\% of the total running requests during each compression operation. Assuming the prefilling length and entry time of each request into the running queue are random, the theoretical proportion of requests needing compression at each step is approximately $\frac{1}{b}$ of the total running requests (where the block size $b$ is 256 in our experiments).

This shows that only a small fraction of requests require compression in each step. If compression and decoding are executed sequentially, many requests that do not require compression will be unnecessarily delayed, waiting for the compression of a few requests to finish. Additionally, the small batch size of compression fails to fully utilize the GPU's computational resources, significantly lowering GPU efficiency.

To resolve this, we enable asynchronous execution of compression and decoding. Requests ready for decoding proceed without waiting for compression to finish, while those requiring compression rejoin subsequent decoding steps once asynchronous compression is complete. This design significantly improves GPU utilization and overall throughput.

\section{Experiments}
\subsection{Experimental Setup}

We conducted experiments using the Qwen3 series models (0.6B, 8B, 14B, and 32B) \citep{yang2025qwen3} and DeepSeek-R1 Distill Llama  8B (referred to as DS Llama 8B) model \citep{guo2025deepseek}.  We adopt an offline inference manner for evaluation. Except for Qwen3 32B, which runs on 2 A100 GPUs using tensor parallelism, all other experiments are conducted on a single A100 GPU. The block size $b$ was fixed at 256, and the window size $w$ was set to 16. We experimented with larger window sizes, the performance showed almost no difference or even worse, but the memory required to store the queries increased significantly.

\begin{figure*}[t]
    \centering
    \includegraphics[width=0.72\linewidth]{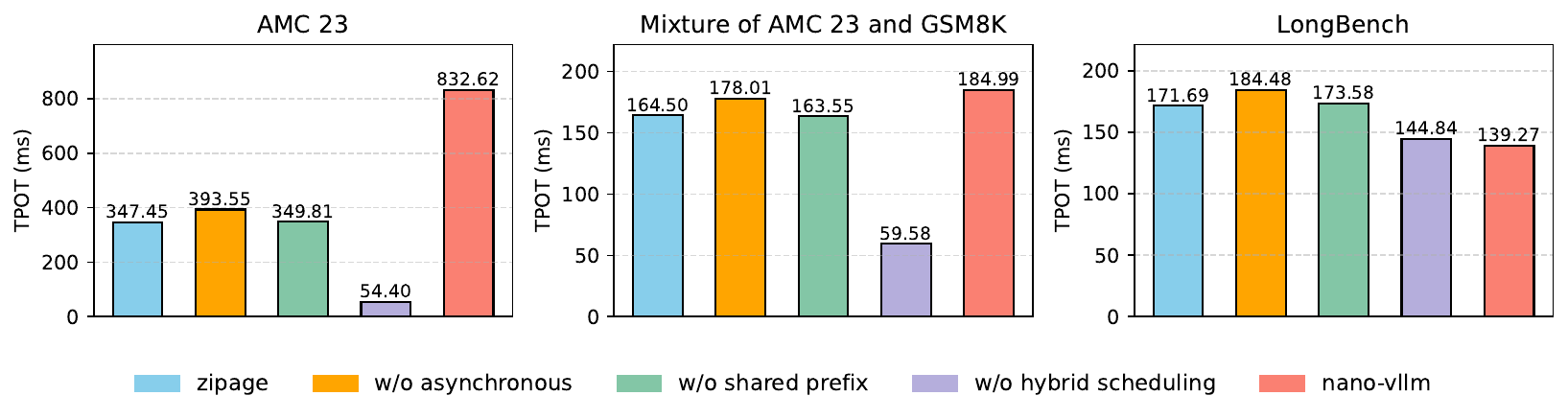}
    \caption{Comparison of average TPOT (ms) of all requests across different configurations on three workloads.}
    \label{fig:ablation_tpot}
\end{figure*}

\begin{figure*}[t]
    \centering
    \includegraphics[width=0.72\linewidth]{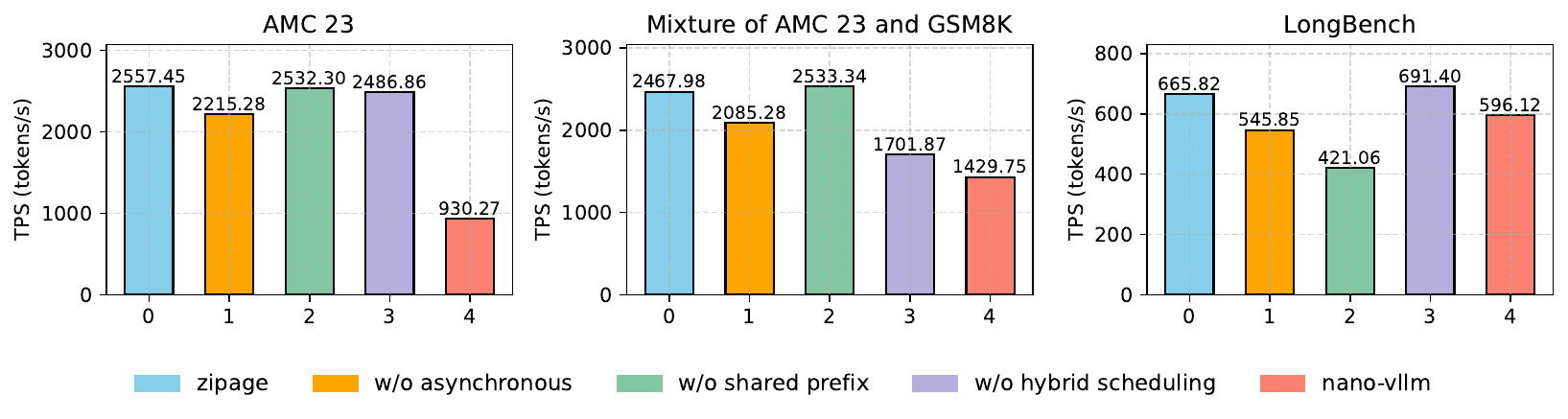}
    \caption{Comparison of TPS (tokens/s) across different configurations on three workloads.}
    \label{fig:ablation_tps}
\end{figure*}

To evaluate efficiency, we use two metrics: time per output token (TPOT) and tokens per second (TPS). TPOT is calculated as the total time from the generation of the first token to the last token for a request, divided by the total number of tokens generated for that request. TPS is defined as the total number of tokens generated across all requests, divided by the total time from the start of the first request to the completion of the last. Model performance is assessed using pass@1 as the evaluation metric \citep{chen2021evaluating}. Evaluation settings and benchmark details are provided in Appendix \ref{sct:eval_data}.

\subsection{Efficiency Analysis}

To evaluate the efficiency of Zipage, we selected three distinct workload types. The first is the mathematical benchmark AMC 23\footnote{https://huggingface.co/datasets/math-ai/amc23}, characterized by short inputs and long outputs. GSM8K\footnote{https://huggingface.co/datasets/openai/gsm8k}, by contrast, is a simpler mathematical benchmark with both short inputs and short outputs. For mixed workloads, we combined AMC 23 and GSM8K. Lastly, we selected the MultiFieldQA task from LongBench \citep{bai2024longbench} as a representative workload with long inputs and short outputs. The KV cache budget fixed at 2048.

In addition to evaluations using Zipage, ablation studies were conducted on three techniques: asynchronous compression, hybrid scheduling, and prefix sharing. Furthermore, comparisons were made with Nano-vLLM\footnote{https://github.com/GeeeekExplorer/nano-vllm}, a lightweight implementation of PagedAttention.

Figure \ref{fig:ablation_tpot} shows the TPOT of Qwen3 8B under different configurations. TPOT decreases significantly when hybrid scheduling is disabled, as requests are rarely preempted or blocked, allowing uninterrupted decoding until completion. In contrast, for Zipage with hybrid scheduling or Nano-vLLM, the re-queuing time after preemption can dominate the overall request processing time.  \textbf{Thus, the TPOT metric becomes less meaningful. Our subsequent analysis will focus primarily on the TPS metric.}

\begin{figure*}[ht]
  \centering

  \begin{subfigure}[t]{0.37\linewidth}
    \centering
    \includegraphics[width=\linewidth]{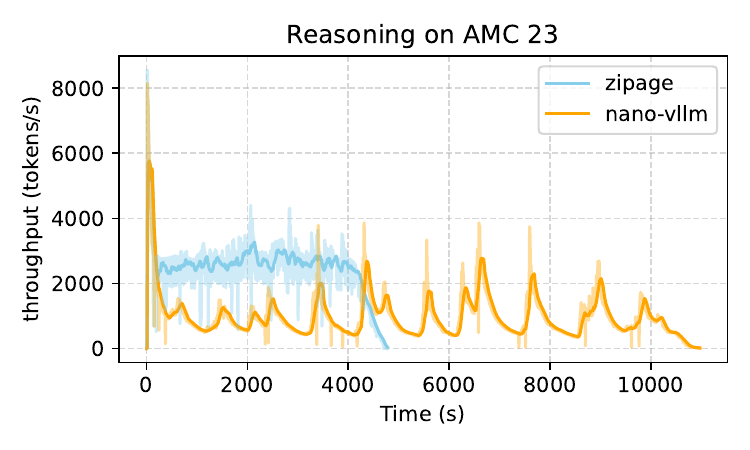}
    \caption{}
  \end{subfigure}
  \begin{subfigure}[t]{0.37\linewidth}
    \centering
    \includegraphics[width=\linewidth]{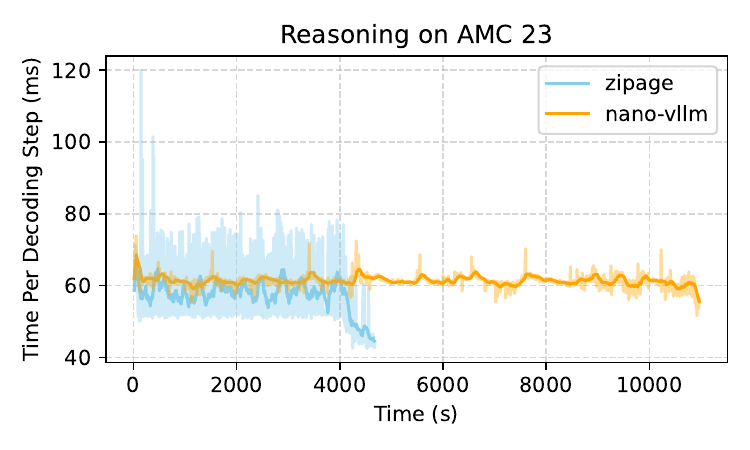}
    \caption{}
  \end{subfigure}

  \begin{subfigure}[t]{0.37\linewidth}
    \centering
    \includegraphics[width=\linewidth, trim=0 0 0 0, clip]{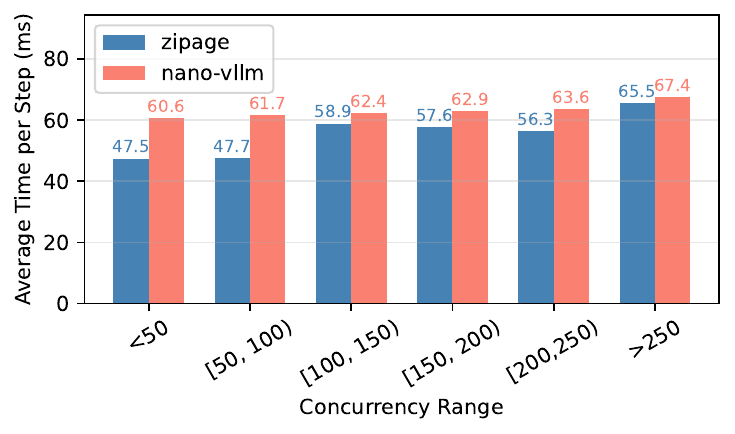}
    \caption{}
  \end{subfigure}
  \begin{subfigure}[t]{0.37\linewidth}
    \centering
    \includegraphics[width=\linewidth, trim=0 0 0 0, clip]{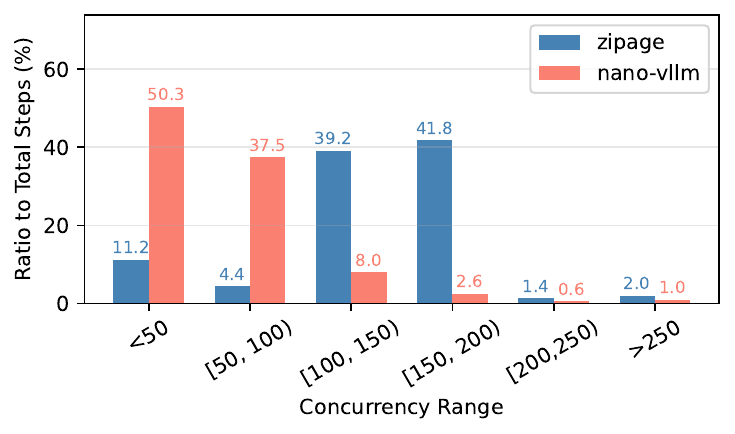}
    \caption{}
  \end{subfigure}

  \caption{The figure shows Qwen3 8B's performance using Zipage or Nano-vLLM on AMC 23, including:(a) real-time throughput, (b) per-step real-time decoding time, (c) average per-step time at different concurrency range, (d) and the ratio of steps to total steps under different concurrency range.}
  \label{fig:real_time}
\end{figure*}

Figure \ref{fig:ablation_tps} shows the TPS of Qwen3 8B under various configurations. Disabling asynchronous compression consistently lowers TPS across all workloads, underscoring its acceleration benefits in all scenarios. Hybrid scheduling proves advantageous in mixed workloads dominated by short-input and short-output requests, as it improves concurrency. Prefix caching significantly speeds up LongBench due to the presence of long shared prefixes in this workload. Zipage outperforms Nano-vLLM in the TPS metric, with its advantage becoming more evident in scenarios requiring longer outputs. Additional details, including the number of running and waiting requests during inference and block utilization rates, are provided in Appendices \ref{sct:ab_info} and \ref{sct:cp_info}.

Figure \ref{fig:real_time} (a) shows the real-time throughput during inference, calculated as the number of tokens decoded per step divided by the decoding time per step. Nano-vLLM exhibits periodic throughput fluctuations due to offloading requests as sequences lengthen. When a long request completes, the offloaded requests rejoin the running queue, temporarily boosting throughput. In contrast, Zipage maintains consistently high throughput, although asynchronous compression, which competes with decoding for GPU resources, introduces some fluctuations.  
Figure \ref{fig:real_time} (b) illustrates the time per step, while Figure \ref{fig:real_time} (c) compares the average execution time per step across different concurrency ranges, showing that Zipage achieves shorter times at the same concurrency levels. Figure \ref{fig:real_time} (d) reveals the proportion of steps across various concurrency ranges, with Zipage primarily operating within the high-concurrency range of [100, 200), whereas Nano-vLLM operates mostly below 100. Additional experimental details for models of different scales are provided in Appendix \ref{sct:scale_info}.

\subsection{Comparison with Other Frameworks}

In this section, we compare Zipage with other text generation frameworks. The baselines include HuggingFace generation\footnote{https://huggingface.co/docs/transformers/index} (HF-Gen) , a Full KV generation framework, as well as MorphKV, R-KV, and G-KV, which incorporate KV cache eviction during decoding to maintain a constant memory. However, none of these methods support advanced techniques such as continuous batching. We also evaluate the inference engines vLLM (v 0.13.0) and Nano-vLLM. vLLM is a highly optimized inference engine for industrial-grade applications.  

\begin{figure}[!ht]
    \centering
    \includegraphics[width=0.99\linewidth]{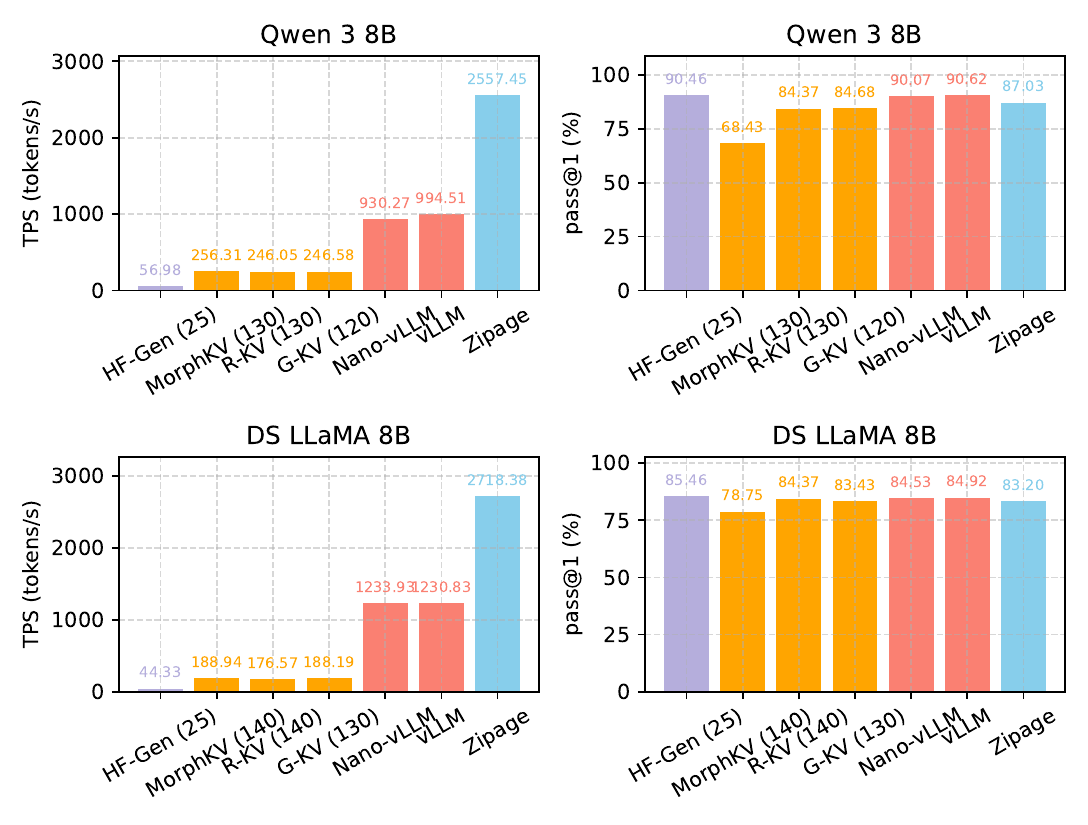}
    \caption{Comparison of TPS and pass@1 performance across different methods for Qwen3 8B and DS LLaMA 8B. The numbers in parentheses indicate the maximum batch size.}
    \label{fig:baseline}
\end{figure}

The evaluation is performed on the AMC 23. For methods that do not support continuous batching, a step size of 5 is used to search for the maximum batch size. For Zipage and other baselines that support KV cache eviction, the KV cache budget is fixed at 2048.
 
\begin{figure*}[!ht]
    \centering
    \begin{subfigure}[t]{0.9\textwidth}
        \centering
        \includegraphics[width=\linewidth]{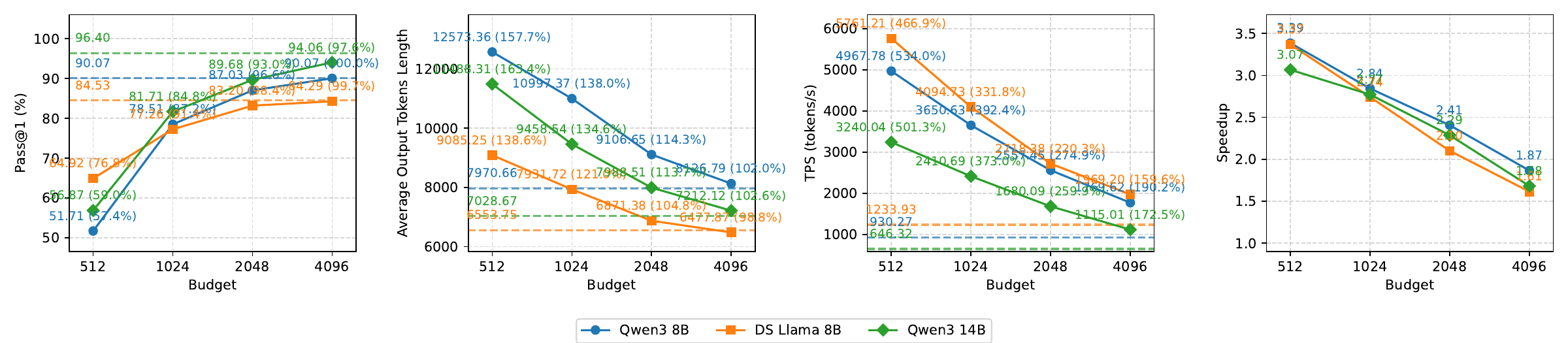}
        \caption{AMC 23}
        \label{fig:amc23}
    \end{subfigure}
    \begin{subfigure}[t]{0.9\textwidth}
        \centering
        \includegraphics[width=\linewidth]{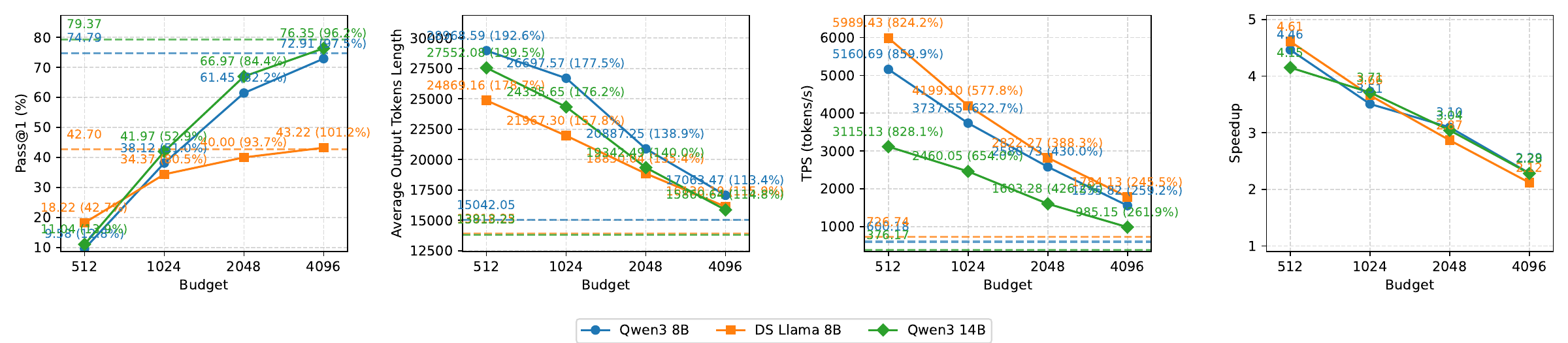}
        \caption{AIME 24}
        \label{fig:aime24}
    \end{subfigure}
    \begin{subfigure}[t]{0.9\textwidth}
        \centering
        \includegraphics[width=\linewidth]{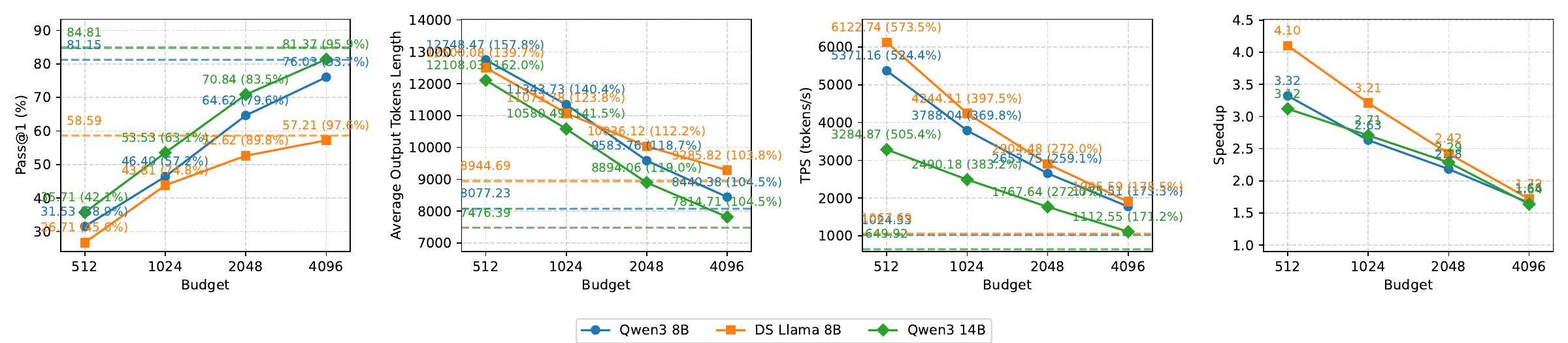}
        \caption{LiveCodeBench}
        \label{fig:code}
    \end{subfigure}
    \caption{Evaluation results under varying KV cache budgets on (a) AMC 23, (b) AIME 24, and (c) LiveCodeBench. Dashed lines represent the results of full KV, and the percentages following the numerical values indicate the relative performance compared to full KV.}
    \label{fig:eval_all}
\end{figure*}

As illustrated in Figure \ref{fig:baseline}, in terms of TPS, methods supporting KV cache eviction demonstrate significant improvements over HF-Gen, primarily due to their capability to handle larger batch sizes. However, these methods, lacking features such as continuous batching, produce a substantial number of padding tokens, causing their TPS to fall below that of inference engines like vLLM and Nano-vLLM. In contrast, Zipage achieves more than double the TPS of both vLLM and Nano-vLLM.  Regarding pass@1 performance, methods utilizing KV cache eviction, including Zipage, deliver results comparable to Full KV cache approaches under a 2k budget, with the exception of MorphKV, which exhibits a slight performance gap.

\subsection{How to Set KV Cache Budgets?}

In this section, we evaluate using different KV cache budgets on two mathematical benchmarks, AMC 23 and AIME 24\footnote{https://huggingface.co/datasets/math-ai/aime24}, and one code benchmark, LiveCodeBench \citep{jain2024livecodebench} v1. We report pass@1, average output length, TPS, and speedup, calculated based on the total time required to complete all requests.

For AMC 23, with a budget of 2048, the performance of Zipage reaches around 95\% of that of Full KV (Nano-vLLM), while throughput and speedup exceed twice the Full KV baseline. At a budget of 4096, the performance is very close to Full KV. For AIME 24, a 4096 budget achieves about 95\% of Full KV performance, with throughput and speedup also exceeding twice the baseline. For code tasks, Zipage achieve around  95\% of Full KV performance with a 4096 budget, with a speedup ratio of approximately 1.6.  Additionally, we observe that as the budget decreases, not only does performance decline, but the average output length also increases, which may negatively affect user experience. 

\section{Discussion}
Zipage currently supports RoPE \citep{su2024roformer} and its variants, which encoding positional embeddings directly into the KV cache. Zipage is fully compatible with text-based context management systems for multi-turn conversations, and integrates seamlessly with FlashAttention \citep{dao2022flashattention} as KV cache eviction does not interfere with the Attention forward process.

\section{Deployment Considerations}

Zipage is designed and evaluated under the assumption of a trusted, single-tenant deployment environment (i.e., a dedicated instance operated by one organization). Accordingly, strong security/privacy isolation between mutually untrusted tenants or workloads is intentionally out of scope for this work. Operators should deploy Zipage within an environment that already provides the required access controls, network boundaries, and compliance safeguards; scenarios requiring multi-tenant isolation or adversarial co-location protections are not addressed here.

\section{Conclusion}

In this paper, we propose Compressed PagedAttention, which integrates KV cache compression with paged KV cache management. Based on this, we develop the inference engine Zipage, which achieves over 2x speedup while delivering performance close to that of Full KV in reasoning tasks.

\section*{Limitations}

Since we have not yet implemented the online engine, all evaluations are conducted in the form of offline inference. Therefore, we do not report the Time to First Token (TTFT) metric, as queuing time dominates and renders TTFT less meaningful as a reference. In the future, we plan to implement an online engine in the open-source project and integrate techniques such as chunked prefilling \citep{agrawal2023sarathi} to optimize TTFT.

We did not compare our approach with methods that integrate KV cache compression into inference engines, as RaaS \cite{raas} and PagedEviction \cite{pagedeviction} lack publicly available code for such integration. However, our token-wise eviction method may offer advantages in preserving critical information. As for KV-Compress \citep{rehg2024kv}, it only compresses inputs and performs similarly to vLLM in scenarios with long outputs.

Additionally, requests of varying difficulty may require different budgets. Currently, we set the budget to a fixed size. However, $N_{\max}$ can be treated as a unique attribute for each request and adjusted based on the actual sequence length of the request, which might slightly reduce concurrency but could improve overall performance. We plan to incorporate this feature into Zipage in the future.


\bibliography{custom}

\appendix

\section{Usage of AI}  
We employed AI to refine the content based on our original text, with all revisions thoroughly reviewed and verified by our team. The code development was assisted with AI. All code underwent rigorous testing.

\section{Evaluation Settings and Dataset Details}
\label{sct:eval_data}

For all evaluations, the sampling temperature is set to 0.6. The evaluation settings for different benchmarks are detailed in Table \ref{tab:eva_setting}.

\begin{table}[!h]
\centering
\resizebox{0.45\textwidth}{!}{%
\begin{tabular}{l c c c}
\toprule
\textbf{Workloads} &
\shortstack{\textbf{Number of} \\ \textbf{Questions}} &
\shortstack{\textbf{Sample} \\ \textbf{Times}} &
\shortstack{\textbf{Max Output} \\ \textbf{Length}} \\
\midrule
AMC 23 & 40 & 32 & 16384 \\
AIME 24 & 30 & 32 & 32768 \\
LiveCodeBench v1 & 400 & 8 & 16384 \\
Mixture & 40+1319 & 4 & 16384 \\
LongBench & 150 & 4 & 4096 \\
\bottomrule
\end{tabular}
}
\caption{Evaluation settings for different workloads or benchmarks.}
\label{tab:eva_setting}
\end{table}

For GSM8K, we use Qwen3's non-reasoning mode, which reduces output length in most cases, though the model occasionally generates lengthy reasoning. Additionally, for the Mixture of GSM8K and AMC23 workload, the question order is randomized.

\section{Implementation for Compression Operations and Experiments.}

For most operations during the compression process, we implemented specialized GPU kernels using Triton\footnote{https://github.com/triton-lang/triton}. This section contains algorithm descriptions and experiments for these kernel implementations.

\subsection{Cross-Layer Parallel Compression}

Since the KV cache compression processes of different layers are independent of each other, compression can be executed in parallel across layers. 
All kernels can parallelize at least across the dimensions of batch size, layer, and attention head.

However, the compression process generates intermediate activations. If all layers are compressed simultaneously, it may lead to memory overflow in extreme cases, especially under asynchronous compression settings where memory is shared with prefilling and decoding operations. To address this, we adapt cross-layer compression based on a layer stride $l$, i.e., compressing the KV cache of $l$ layers at each time. The peak of the activations scales as $\mathcal{O}(n \times l \times h_\text{q} \times N \times b \times w)$, where $n$ is the batch size of requests requiring compression, and $N$ is the maximum number of blocks among all requests.

\begin{figure}[h]
    \centering
    \includegraphics[width=0.95\linewidth]{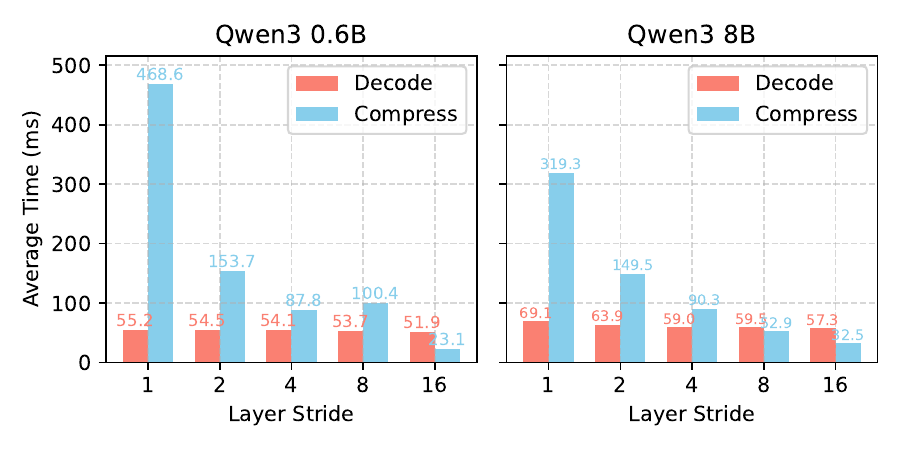}
    \caption{Average time for compression and decoding step under different layer stride.}
    \label{fig:layer_stride}
\end{figure}

We inference on AMC 23 and report the average compression and decoding times for different layer strides are shown in Figure \ref{fig:layer_stride}. As the layer stride increases, both the average decoding time per step and the compression time exhibit a decreasing trend. Notably, the compression time for a layer stride of 16 is approximately $5\%-10\%$ of that for a layer stride of 1. 

Furthermore, we observe that the average compression time for the 0.6B model is longer than that for the 8B model. This is attributed to the higher concurrency in the 0.6B model, resulting in a larger average batch size for each compression step.

Ultimately, we adopt a layer stride of 8 in other experiments, which provides a significant acceleration while maintaining a moderate size for intermediate activation values.

\subsection{Paged Attention Score}

The scoring function $\phi(\mathbf Q, \mathbf K, \mathcal I)$ in its most basic form involves only the computation of attention scores. In this section, we describe how to calculate attention scores using the query states in the query cache $\mathbf Q$ and the key states in the key cache $\mathbf K$. Since the computations for different requests within a batch, different layers, and different attention heads are independent and executed in parallel, \textbf{we illustrate the algorithm with a single request, a single layer, and a single attention head.}

\label{sct:attention_score}
\begin{algorithm}
\caption{Block Attention Logits Computation}
\label{alg:triton-attn-score}
\begin{algorithmic}[1]
\Require 
    Query cache $\mathbf{Q}$, 
    key cache $\mathbf{K}$,
    block size $b$,  
    block index $i$,
    query slots index $j$,
    block table,
    attention dimension $d$
    
\State Compute query offset $p_q$ in $\mathbf{Q}$ based on query slots index $j$
\State Load query states: $\mathbf{Q}_j \gets \mathbf{Q}[p_q] \in \mathbb{R}^{w \times d}$
\State Get the block id in block table through block index $i$
\State Compute key offset $p_k$ in $\mathbf{K}$ using block id
\State Load key states: $\mathbf{K}_i \gets \mathbf{K}[p_k] \in \mathbb{R}^{b \times d}$

\State Compute attention logits: $\mathbf{A'} \gets  \frac{ \mathbf{Q}_j \cdot \mathbf{K}_i^\top}{\sqrt{d}} \in \mathbb{R}^{w \times b}$

\If{block $i$ is the last block the request}
    \State Construct causal mask $\mathbf{M} \in \mathbb{R}^{w \times b}$ where
    \[
    \mathbf M_{u,v} = 
    \begin{cases}
    -\infty & \text{if } u + b - w> v  \\
    0 & \text{otherwise}
    \end{cases}
    \]
    \State Apply mask: $\mathbf{A'} \gets \mathbf{A'} + \mathbf{M}$
\EndIf
\State Save: $\mathbf{A}[i] \gets \mathbf{A'}$
\end{algorithmic}
\end{algorithm}

When computing the attention scores, we first allocate storage space $A \in \mathbb{R}^{N \times w \times b}$. Then, as described in Algorithm \ref{alg:triton-attn-score}, we perform the matrix multiplication of query states and key states. Since the attention computation can be further parallelized along the dimension of block numbers, Algorithm \ref{alg:triton-attn-score} outlines the computation for a single block. 

It is important to note that, although the matrix multiplication in algorithms is conceptually completed in a single step, in practice, it is further divided into smaller blocks for computation. What's more, \textit{load} in algorithms refers to reading data from the GPU's global memory (Dynamic Random Access Memory, DRAM) into the shared memory or registers  (Static Random Access Memory, SRAM), while \textit{save} refers to writing data from the shared memory back to the global memory.

The layout of key states in $\mathbf{K}$ is paged, but the computed logits $\mathbf{A}$ are stored contiguously. So, we can reshape $\mathbf{A}$ into $\mathbb{R}^{w \times (N b)}$ and apply Softmax along the last dimension to obtain the attention scores $\mathbf{S}' \in \mathbb{R}^{w \times (N b)}$. 

For multi-query \citep{shazeer2019fast} or group-query \citep{ainslie2023gqa} attention, where a single key head corresponds to multiple query heads, we perform a max-reduce operation for the scores of each key head. Finally, we take an average along the observation window length dimension to obtain $\mathbf{S} \in \mathbb{R}^{N \times b}$.

\subsection{Global Score}
\label{sct:global_score}

The global score proposed in G-KV \citep{gkv} combines historical attention scores through decayed max-reduce or sum-reduce, enabling better evaluation of the long-term importance of KV cache entries during eviction. We have integrated the global score into our framework and adapted it for PagedAttention.

First, we need to pre-allocate $\mathbf F \in \mathbb{R}^{L \times N_{\text{total}} \times b \times h_{\text{kv}}}$ to store the global score. The size of $\mathbf F$ is $\frac{1}{2d}$ of the total size of $\mathbf{K}$ and $\mathbf{V}$. If global score is enabled, equation (\ref{eq:max_c}) needs be  updated as:
\begin{equation}  
\left\{
\begin{aligned}
    &(1+\frac{1}{2d})\times m_{\text{kv}} \times N_{\text{total}} + M \times m_{\text{q}} \leq m_{\text{available}}, \\
    &M \leq \frac{N_{\text{total}}}{N_{\max}}, \\
    &M > 0, \quad N_{\text{total}} > 0,
\end{aligned}
\right.
\end{equation}

\begin{algorithm}
\caption{Update Global Score Cache with Attention Scores}
\label{alg:global_score}
\begin{algorithmic}[1]
\Require Attention score tensor $\mathbf S$, global score cache $\mathbf F$, block table, number of blocks $N$, decay factor $\alpha$, block index $i$

\State Load attention scores of $i$-th block  $\mathbf s_i \in \mathbb{R}^b$ from $\mathbf S$
\State Compute offset $p$ of $i$-th block  in $\mathbf F$ using block id from block table
\If{request is not compressed}
    \State Save score to cache: $ \mathbf F[p] \gets \mathbf s_i$
\Else
    \If{$i$-th block  is not the last block of the sequence}
        \State Load previous global score $\mathbf f_i \gets \mathbf F[p]\in \mathbb{R}^b$
        \State Update: $\mathbf s_i \gets  \max(\alpha \cdot\mathbf f_i , \mathbf s_i)$
    \EndIf
    \State Save score to cache: $ \mathbf F[p] \gets \mathbf s_i$
    \State Overwrite attention score: $\mathbf S[i] \gets\mathbf s_i$
\EndIf
\end{algorithmic}
\end{algorithm}

Based on the attention scores $\mathbf{S}$, we use Algorithm \ref{alg:global_score} to compute the global scores. The algorithm can be summarized as follows: if a request has not been compressed, there are no historical scores, and we simply store $\mathbf{S}$ in $\mathbf{F}$. If a request has been compressed, all blocks except the last one have historical scores. For these blocks, we take the maximum value between the decayed historical scores and $\mathbf{S}$ as the new score.

\begin{table}[h]
\centering
\resizebox{\linewidth}{!}{%
\begin{tabular}{c|cccccc}
\toprule
$\alpha$ & 0 & 0.4 & 0.8 & 0.85 & 0.9 &1\\
\midrule
Qwen3 8B      & 0.6906 & 0.7375 & \bfseries 0.7718 & 0.7468 & 0.7484 & 0.7531 \\
DS Llama 8B   & 0.7484 & \bfseries 0.7656 & 0.7643 & 0.7584 & 0.7578 & 0.7515 \\
\bottomrule
\end{tabular}%
}
\caption{ Experimental results with different decay rates $\alpha$.}
\label{tab:global_score}
\end{table}

We evaluate using a budget of 2048 and global score with different decay rates $\alpha$ on the AMC 23 benchmark, with the experimental results shown in Table \ref{tab:global_score}. The global score shows a significant improvement on Qwen3 8B but provides minimal benefits on DS Llama 8B.

\subsection{Pooling at the Sequence Dimension}

SnapKV \citep{li2024snapkv} performs max pooling along the sequence dimension, meaning that tokens near high-scoring tokens are also assigned high scores. This helps preserve more detailed information in context. We have integrated this method into our framework, implementing it using PyTorch\footnote{https://pytorch.org/}'s  MaxPool1D interfaces instead of a specialized kernel:
\begin{equation}
    \mathbf{S}=\text{MaxPool1D}(\mathbf{S})
\end{equation}

Pooling is performed after computing the global score during the  compression process.

\subsection{Redundancy Score of Key states}
\label{sct:similarity_score}

\begin{figure}[ht]
    \centering
    \includegraphics[width=0.98\linewidth, trim=240 20 240 20, clip]{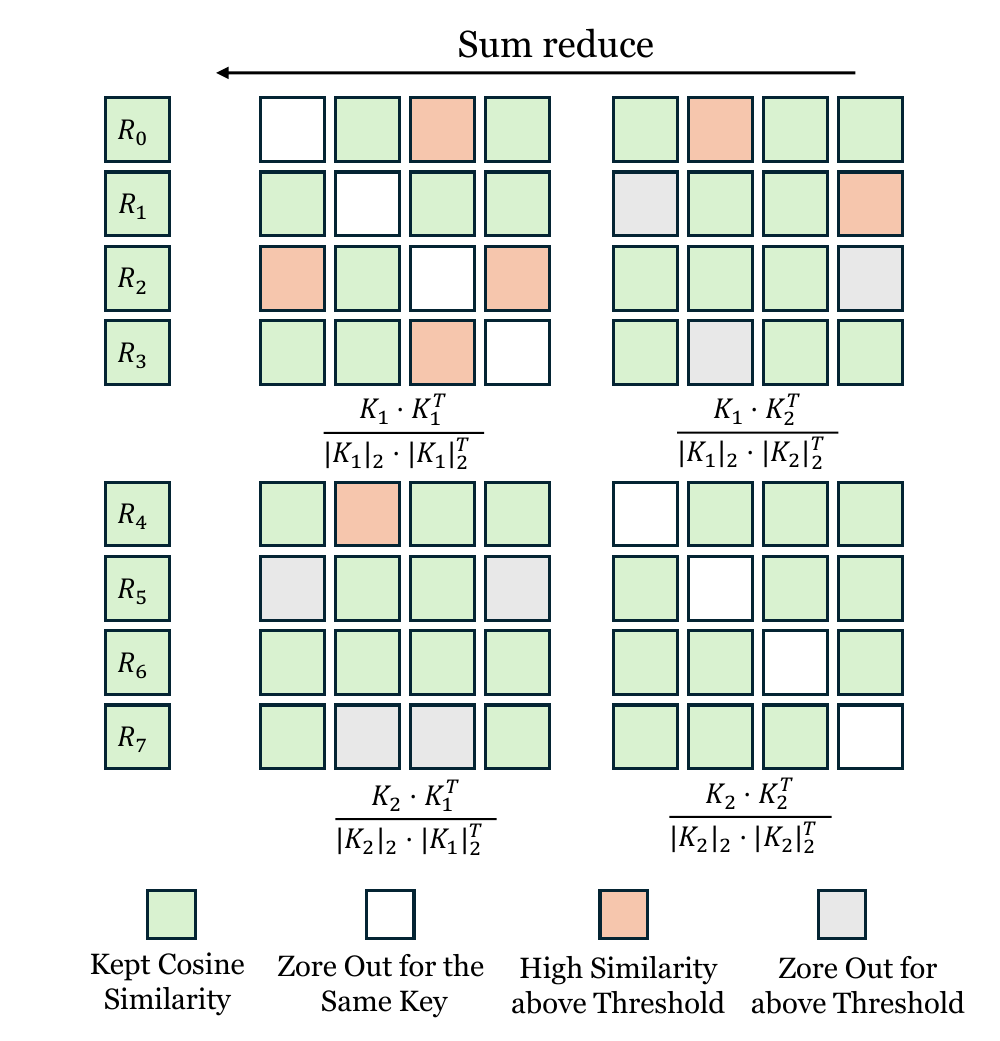}
    \caption{This figure illustrates the computation of the original redundancy scores when $N=2$ and $b=4$. Here, $\mathbf{K}_i$ represents the key states of the $i$-th block. This approach has a computational complexity of $\mathcal{O}(N^2 \times b^2)$ and a memory complexity of $\mathcal{O}(N^2 \times b^2)$.}
    \label{fig:raw_redundancy}
\end{figure}

R-KV \citep{rkv} introduces redundancy scores to evaluate the degree of redundancy among KV cache entries. Specifically, it calculates the cosine similarity between the key states within a sequence. 

Figure \ref{fig:raw_redundancy} illustrates the computation of redundancy scores between the key states of two blocks. The diagonal entries represent the similarity of key states with themselves and are therefore zeroed out. Additionally, for each column, the last similarity score exceeding the threshold $p$ is set to 0, as we prioritize retaining newer tokens when an old token is highly similar to a new token. Finally, the similarity matrix is summed row-wise, normalized by the sequence length, and passed through Softmax on the sequence dimension to compute the redundancy score $\mathbf{R}$. 

The redundancy score is applied after max pooling. The redundancy score is combined with previous scores using the following formula:
\begin{equation}
    \mathbf S=\mathbf{S} - \lambda \cdot\mathbf R
\end{equation}

\begin{algorithm}[!ht]
\caption{Flash Redundancy Score}
\label{alg:redundancy}
\begin{algorithmic}[1]
\Require Key cache $\mathbf K$, block table, number of blocks $N$, threshold $p$, block size $b$, block index $m$
\Ensure Accumulated similarity score $\mathbf R'$
\State Calculate the offset $p_m$ in $\mathbf{K}$ based on the block id of the $m$-th block
\State Load key states $\mathbf K_m  \leftarrow \mathbf{K}[p_m]\in \mathbb R^{b \times d}$ of block $m$
\State Initialize zero-out tag $\mathbf z \in \mathbb R^{1 \times b} \gets 0$
\For{$i = N-1$ to $0$}
    \State Calculate the offset $p_i$ in $\mathbf{K}$ based on the block id of the $i$-th block
    \State Load key states $\mathbf K_i  \leftarrow \mathbf{K}[p_i]\in \mathbb R^{b \times d}$ of block $i$
    \State Compute cosine similarity: $\mathbf C = \frac{\mathbf K_i\cdot\mathbf  K_m^\top}{\|\mathbf K_i\|_2 \cdot \| \mathbf K_m\|_2^T} \in \mathbb R^{b \times b}$
    \If{$i = m$}
        \State Mask diagonal of $\mathbf C \gets 0$
    \EndIf
    \State Identify the last element \(> p\) in the column of \(\mathbf C\), ensure that the corresponding tag in \(\mathbf{z}\) for this column is \(0\), and set this element to \(0\)
    \State Update the tag corresponding in $\mathbf z_m$ to 1 where such element were zero out
    \State  $\mathbf C' \in \mathbb R^{b\times 1} \leftarrow$ Row-wise accumulate $\mathbf C $ 
    \State save the result to $\mathbf R'[i,m] \leftarrow \mathbf C'$
\EndFor
\end{algorithmic}
\end{algorithm}

The original implementation of the redundancy score first computes the complete cosine similarity matrix and then applies zeroing out. Its memory complexity is $\mathcal{O}(N^2 \times b^2)$. Assuming a floating-point size of 2 bytes, the actual matrix size is $2 \times n \times l \times h_\text{kv} \times N^2 \times b^2$. For a common scenario where $n=16$, $l=8$, $h_\text{kv}=8$, and $b=256$, the cosine similarity matrix size is $128 \times N^2$ MB. When the sequence length is sufficiently large, the memory usage can even reach tens of GB.

\begin{figure*}[ht]
    \centering
    \includegraphics[width=0.75\linewidth, trim=80 10 20 10, clip]{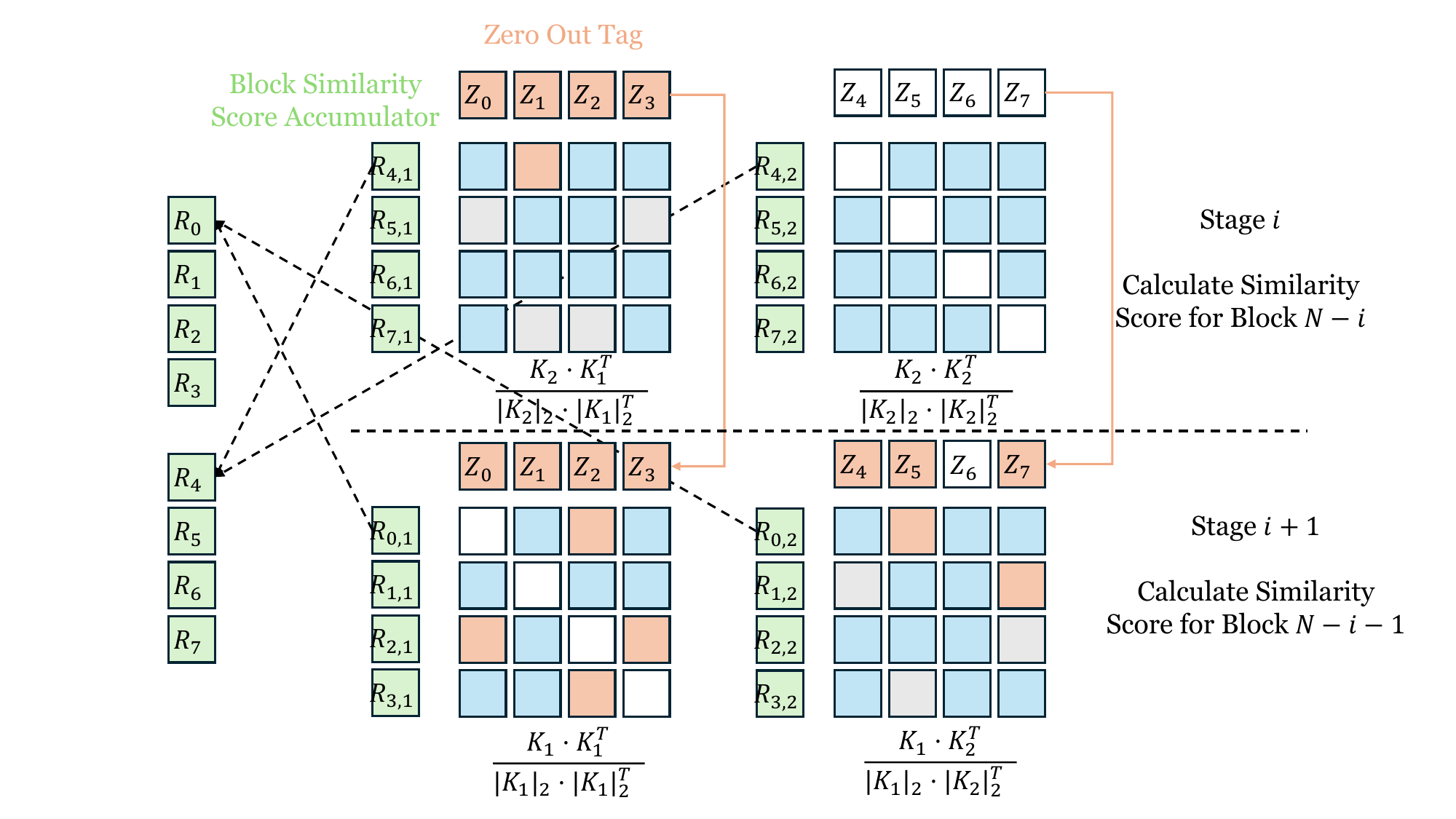}
    \caption{This figure illustrates the computation of flash redundancy score. The green blocks represent data stored in global memory, i.e., activations. The results of other computations, shown in different colors, are temporarily stored using registers or shared memory. This approach has a computational complexity of $\mathcal{O}(N^2 \times b^2)$ and a memory complexity of $\mathcal{O}(N^2 \times b)$.}
    \label{fig:flahs_redundancy}
\end{figure*}

Such enormous activations are unacceptable and can easily lead to memory overflow. To address this issue, we implemented the flash redundancy score. Figure \ref{fig:flahs_redundancy} illustrates the computation process of the flash redundancy score. We no longer store the complete similarity matrix. Instead, we compute similarities in a block-wise manner, starting from the last block. The computed similarity results are not retained but are directly accumulated into a pre-allocated accumulation  accumulator $\mathbf{R}' \in \mathbb{R}^{N \times N \times b}$. To correctly zero out the last high-threshold similarity in each column, we maintain a zero-out tag to track whether the last similarity score exceeding the threshold in each column has been zeroed out. If a column has not been zeroed out, the last value in the block that exceeds the threshold is set to 0, and the corresponding zero-out tag will be set. The detailed process is shown in Algorithm \ref{alg:redundancy}. Finally, we perform sequential accumulation to obtain $\sum_{m=0}^{N-1} \mathbf R'[:,m]$, and then apply length normalization and Softmax to compute $\mathbf{R}$.

The flash redundancy score reduces the memory complexity to $\mathcal{O}(N^2 \times b)$. By partitioning the matrix multiplication into smaller blocks, the intermediate similarity scores and zero-out tags are temporarily stored using registers and shared memory. Only the accumulated results need to be written to $\mathbf{R}'$ in the global memory. In the previous example, the activation size of the flash redundancy score is approximately $\frac{N^2}{2}$ MB, which is $\frac{1}{256}$ of the original implementation.

An even more aggressive implementation exists, where the accumulated activations for similarity scores only require $\mathcal{O}(N \times b)$ memory space. This approach involves directly accumulating the results of $\frac{\mathbf{K}_i \cdot (\mathbf{K}_1;\dots;\mathbf{K}_N)^\top}{|\mathbf{K}_i|_2 \cdot |\mathbf{K}_1;\dots;\mathbf{K}_N|_2^\top} \in \mathbb{R}^{b \times (bN)}$. However, due to the limited capacity of registers and shared memory, only small-scale matrix operations, such as $16 \times 64$, can be performed at a time. Large matrix computations require the kernel to execute additional iterations, which reduces the overall level of parallelism. Our current implementation adopts a balanced approach, trading off between memory usage and parallelism.

\subsection{KV Cache Compression}
\label{sct:compression}

\begin{algorithm}[!h]
\caption{KV Cache Compression}
\label{alg:compress}
\label{alg:kv-compression}
\begin{algorithmic}[1]
\Require Key and Value cache tensors $\mathbf{K}, \mathbf{V} $, block size $b$, number of blocks $N$, top-$k$ tag $\mathbf{F} \in \{0, 1\}^{N \times b}$
\Ensure Compressed $\mathbf{K}$, $\mathbf{V}$

\State Initialize read offset $p_{r}$ and write offset $p_{w}$ to the first slot of the first block
\State $\ell \gets 0$, $s \gets 0$, $i \gets 0$

\While{$i < N$}
    \If{$\mathbf{F}[i][\ell \bmod b] = 1$}
        \State Load key vector $\mathbf{k} \leftarrow\mathbf{ K}[p_{r}] \in \mathbb{R}^{1 \times d}$
        \State Load value vector $\mathbf{v} \leftarrow \mathbf{V}[p_{r}] \in \mathbb{R}^{1 \times d}$
        \State Store $\mathbf{k}$ to $\mathbf{K}[p_{w}]$
        \State Store $\mathbf{v}$ to $\mathbf{V}[p_{w}]$
        \State $s \gets s + 1$
        \If{$s \bmod b = 0$}
            \State Move $p_{w}$ to the first slot of the next block
        \Else
            \State Increase $p_{w}$ to the next slot
        \EndIf
    \EndIf

    \State $\ell \gets \ell + 1$
    \If{$\ell \bmod b = 0$}
        \State Move $p_{r}$ to the first slot of the next block
        \State $i \gets i + 1$
    \Else
        \State Increase $p_{r}$ to the next slot
    \EndIf
\EndWhile
\end{algorithmic}
\end{algorithm}

After obtaining the final scores, we first set the scores corresponding to the observation window to $+\infty$. The kernel implementation for this step is relatively straightforward and will not be discussed in detail. Subsequently, we generate a top-$k$ tag $\mathbf{T} \in \mathbb{R}^{N \times b}$ ($k=(N_{\max}-1)\times b$), where the tag of $k$ KV cache entries with the highest scores along the sequence dimension are marked as 1, while the remaining entries are marked as 0. The top-$k$ tagging is implemented using PyTorch's built-in interfaces. Based on the top-$k$ tag, we reorganize the KV cache placement to ensure that the retained KV cache entries are densely packed in memory.

The algorithm for the compression process is shown in Algorithm \ref{alg:compress}. Simply put, it is based on the movement of data in memory using two pointers. In total, it requires $(N_{\max} - 1) \times b$ reads and writes to the KV cache.

It should be noted that when using the global score, the historical scores stored in $\mathbf{F}$ also need to be moved correspondingly to correctly match the associated KV cache entries. The process is similar to Algorithm \ref{alg:compress}, but the amount of data moved each time is reduced from $d$ to $1$.

\subsection{Lightning Redundancy}
\label{sct:lightning_similarity}

\begin{figure}[ht]
    \centering
    \includegraphics[width=0.95\linewidth]{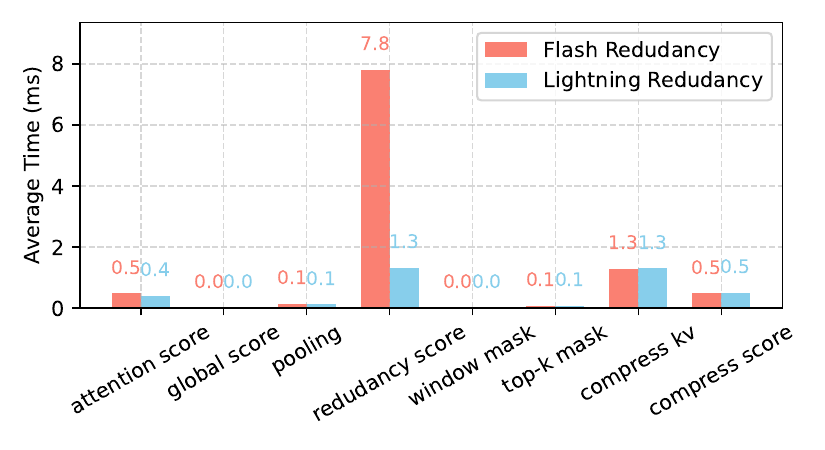}
    \caption{The average execution time of different operations when asynchronous compression is disenabled. A value of 0.0 indicates that the average execution time is less than 0.1 milliseconds.}
    \label{fig:kernel_time_nas}
\end{figure}

\begin{figure}[ht]
    \centering
    \includegraphics[width=0.95\linewidth]{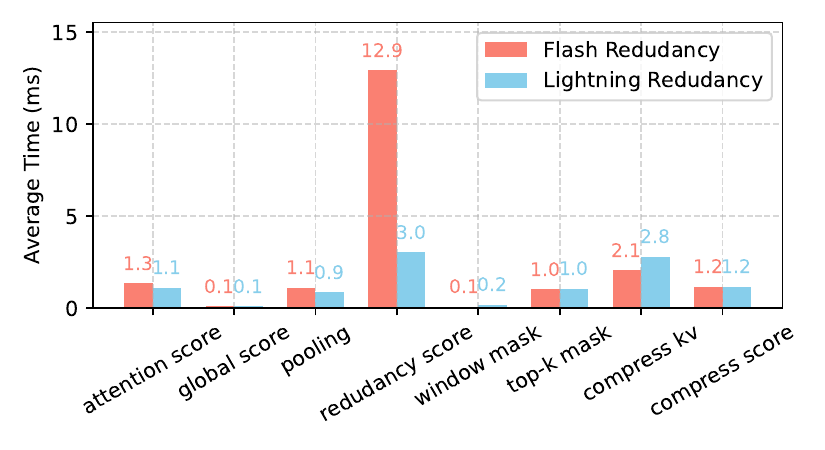}
    \caption{The average execution time of different operations when asynchronous compression is enabled.}
    \label{fig:kernel_time_as}
\end{figure}

The previous sections have introduced all the operations involved in the compression process. We visualized the average execution time of each operation during inference with Qwen3 8B on the AMC 23 benchmark, as shown in Figure \ref{fig:kernel_time_nas} (non-asynchronous compression) and Figure \ref{fig:kernel_time_as} (asynchronous compression), with the red bars representing the execution time. It is evident that the computation of redundancy scores is the bottleneck in the compression process, requiring 1-2 orders of magnitude more time than other operations. This is due to its computational complexity of $\mathcal{O}(N^2 \times b^2)$, which is nearly equivalent to the complexity of attention computation during prefilling.

\begin{figure}[ht]
    \centering
    \includegraphics[width=0.8\linewidth, trim=280 140 240 140, clip]{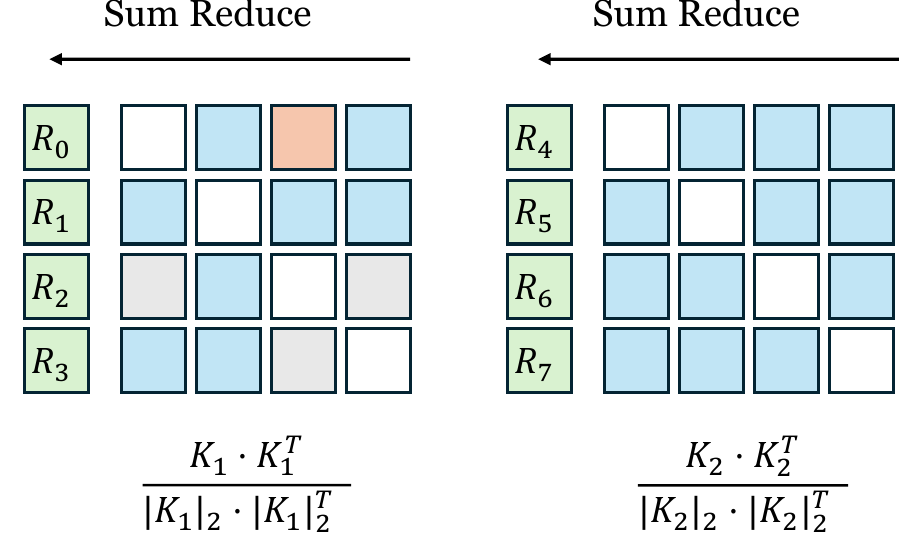}
    \caption{This figure illustrates the computation of lightning redundancy score. This approach has a computational complexity of $\mathcal{O}(N \times b^2)$ and a memory complexity of $\mathcal{O}(N \times b)$.}
    \label{fig:lightning_view}
\end{figure}

To accelerate the compression process, we propose a novel  \textbf{lightning redundancy score}. Specifically, since highly similar representations exhibit locality in the sequence space \citep{lee2025shared}, meaning that the hidden representation of a token is more similar to those of nearby tokens, this phenomenon may be attributed to the attention mechanism focusing more on tokens in close proximity \citep{tan2024stick, chen2024core}. Based on this observation, we propose computing similarity only between keys within the same block and zeroing out only the last similarity score in each column that exceeds the threshold within the block. We refer to this approach as the lightning redundancy score, and Figure \ref{fig:lightning_view} illustrates its calculation process.

\begin{figure}[ht]
    \centering
    \includegraphics[width=0.85\linewidth]{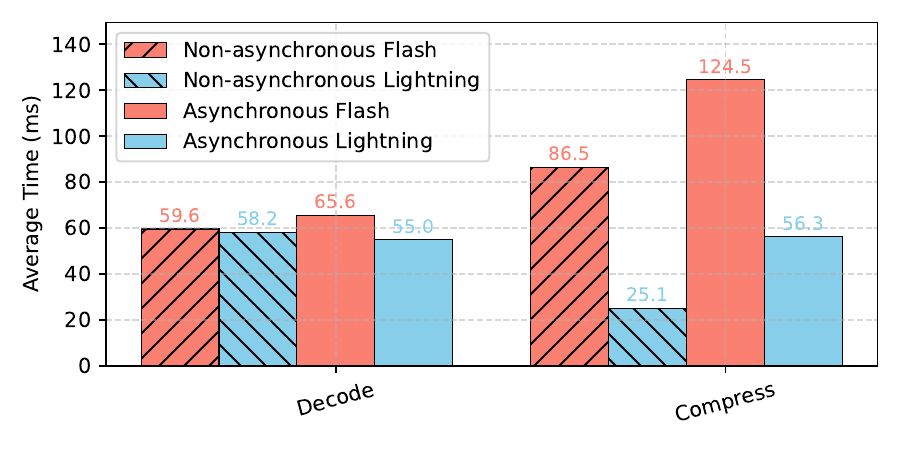}
    \caption{The average decoding time and compression time under both asynchronous and non-asynchronous compression settings.}
    \label{fig:lightning_stage_time}
\end{figure}

The lightning redundancy score reduces the computational complexity to $\mathcal{O}(N \times b^2)$ and the memory complexity to $\mathcal{O}(N \times b)$. As illustrated by the blue bars in Figure \ref{fig:kernel_time_nas} and Figure \ref{fig:kernel_time_as}, the computation time for the lightning redundancy score is significantly reduced. Figure \ref{fig:lightning_stage_time} further presents the average decoding time and compression time under both asynchronous and non-asynchronous compression settings. It is evident that the lightning redundancy score decreases the compression time to a level comparable to that of single-step decoding, without impacting the asynchronously executed decoding process. In contrast, the flash redundancy score, due to its substantial computational overhead, intensifies resource contention with the decoding threads, leading to an increase in average decoding time.

\begin{table}[h]
\centering
\resizebox{\linewidth}{!}{%
\begin{tabular}{c|ccccc}
\toprule
$\lambda$ & 0.05 & 0.1 & 0.2 & 0.5 & 0.9 \\
\midrule
Flash      & 80.78     & 82.96     & 84.37     & \textbf{84.53} & \textbf{77.81} \\
Lightning  & \textbf{83.59} & \textbf{84.84} & \textbf{84.84} & 84.21     & 75.00     \\
\bottomrule
\end{tabular}%
}
\caption{Qwen3 8B on AMC 23}
\label{tab:flc_qwen}
\end{table}

\begin{table}[h]
\centering
\resizebox{\linewidth}{!}{%
\begin{tabular}{c|ccccc}
\toprule
$\lambda$ & 0.05 & 0.1 & 0.2 & 0.5 & 0.9 \\
\midrule
Flash     & \textbf{82.96} & 82.03 & 83.59 & 78.59 & \textbf{74.21} \\
Lightning & 80.62 & \textbf{85.15} & \textbf{85.00} & \textbf{80.78} & 73.90 \\
\bottomrule
\end{tabular}%
}
\caption{DS-Llama 8B on AMC 23}
\label{tab:flc_llama}
\end{table}

Finally, we compare the performance of the two different redundancy scores under various hyperparameters $\lambda$. As shown in Table \ref{tab:flc_qwen} and Table \ref{tab:flc_llama}, in most cases, the lightning redundancy score achieves even better performance.

\subsection{Combining All These Techniques}

In this section, we aim to combine the previously discussed methods. G-KV \citep{gkv} has attempted to integrate the global score and the redundancy score. However, due to the use of max normalization for the scores, it required re-tuning the hyperparameter $\lambda$, making it more sensitive to parameter selection. In our approach, we eliminate the max normalization. Additionally, the redundancy score is ultimately calculated as a distribution via softmax. We observed that this distribution is relatively uniform, especially in the shallow layers, where the differences between scores are minimal. To address this, we introduce a temperature parameter $\tau$ for the softmax computation of the redundancy score, which amplifies highly redundant scores.

\begin{figure}[h]
    \centering
    \includegraphics[width=1\linewidth]{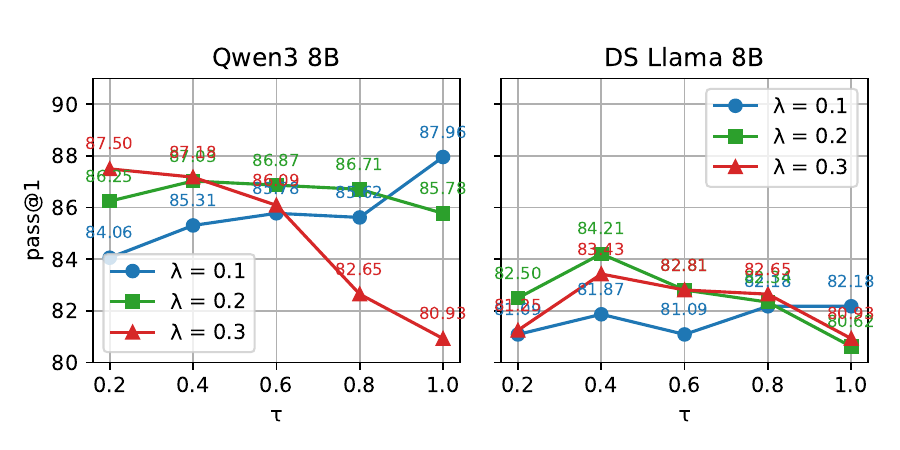}
    \caption{ pass@1 performance for Qwen3 8B and DS Llama 8B under diffirent $\lambda$  and $\tau$.}
    \label{fig:temp_exp}
\end{figure}

For the joint search of $\lambda$ and $\tau$ ($\alpha=0.8$ for global score), the experimental results are shown in Figure \ref{fig:temp_exp}. The temperature has a significant impact on Qwen3 8B; as $\lambda$ increases, representing a higher proportion of the global score, lowering the temperature often yields better results. However, for DS Llama 8B, the final performance is less sensitive to the relationship between $\lambda$ and $\tau$.

\begin{figure}[h]
    \centering
    \includegraphics[width=1\linewidth]{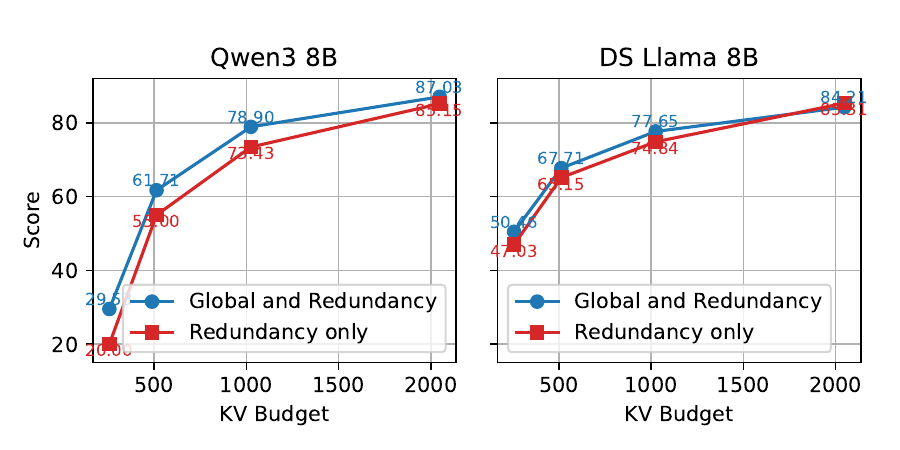}
    \caption{Comparison under different KV budgets for Qwen3 8B and DS Llama 8B, using global and redundancy scores versus redundancy-only scores.}
    \label{fig:combine_exp}
\end{figure}

Figure \ref{fig:combine_exp} presents the ablation study on the use of the global score. For the Qwen3 8B model, the global score provides significant benefits. However, for DS Llama 8B, when the budget is sufficient (2048), not using the global score can even yield better results.

\begin{table}[ht]
\centering
\label{tab:pooling-strategies}
\resizebox{\linewidth}{!}{%
\begin{tabular}{lccc}
\toprule
Model & {w/o pooling} & {pooling once} & {w/ pooling} \\
\midrule
Qwen3 8B  & \textbf{87.10} & 87.03 & 80.85 \\
DS Llama 8B & 84.21 & 83.20 & \textbf{84.53} \\
\bottomrule
\end{tabular}
}
\caption{Comparison of pooling strategies for LLaMA-8B and Qwen-8B.}
\label{fig:pooling}
\end{table}

Finally, we evaluate the impact of max pooling. We use three settings: no pooling at all, pooling only during the first compression, and pooling during every compression step. As shown in Figure \ref{fig:pooling}, for DS Llama 8B, the performance across all settings is relatively similar. However, for Qwen3 8B, performing pooling at every compression step leads to a significant drop in performance.

Although pooling  does not show significant improvements in this scenario, it has proven to be effective in settings based solely on attention scores. The effectiveness of pooling may stem from its ability to retain some important tokens that are temporarily not attended to by the observation window. The global score serves a similar purpose, but demonstrates better performance. Introducing pooling alongside the global score may, in fact, hinder the eviction of less important KV cache entries. However, since the first compression step lacks global scores, pooling can still play a useful role. Therefore, we recommend using  pooling at first compression.

Based on the aforementioned results, we recommend the configuration of $\lambda=0.2$, $\tau=0.4$, and $\alpha=0.8$, with pooling applied only during the first compression step. Although these suggestions are derived from evaluations on a single dataset and therefore have limited generalizability, they have already achieved relatively optimal performance.

\section{Additional Information of Ablation Experiments}

\label{sct:ab_info}

In this section, we provide additional information during the inference process, such as the number of running requests, the number of waiting requests, and block utilization rates.

\begin{figure}[ht]
  \centering
  \begin{subfigure}[t]{0.74\linewidth}
    \centering
    \includegraphics[width=\linewidth, ]{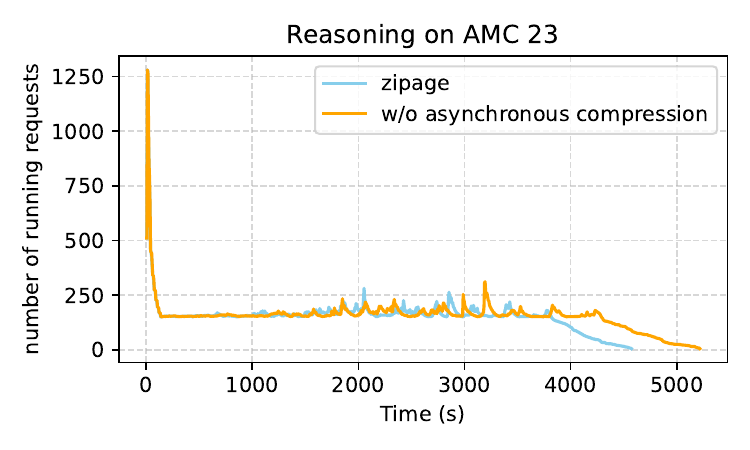}
    \caption{}
  \end{subfigure}

  \begin{subfigure}[t]{0.74\linewidth}
    \centering
    \includegraphics[width=\linewidth]{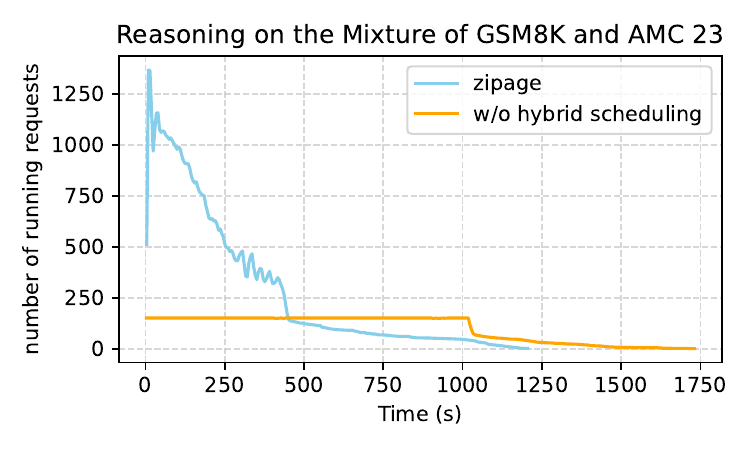}
    \caption{}
  \end{subfigure}
  \begin{subfigure}[t]{0.74\linewidth}
    \centering
    \includegraphics[width=\linewidth]{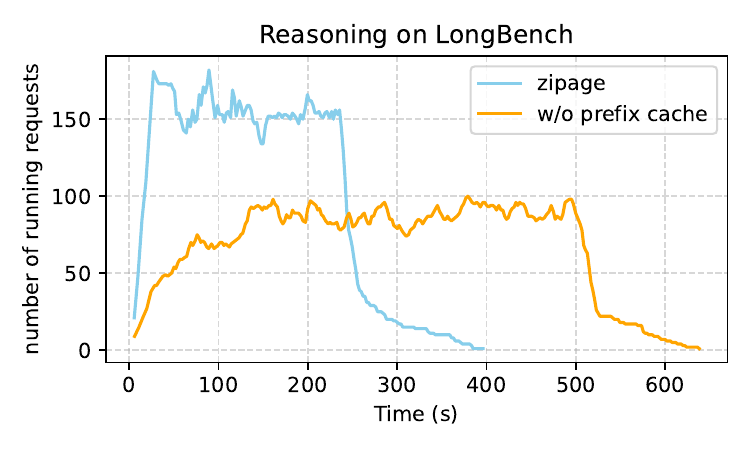}
    \caption{}
  \end{subfigure}

  \caption{Number of running requests during inference.}
  \label{fig:running_ab}
\end{figure}

Figure \ref{fig:running_ab} illustrates the number of running requests during inference. As shown in Figure \ref{fig:running_ab} (b), without hybrid scheduling, the number of running requests remains at or below the maximum concurrency. In contrast, with hybrid scheduling enabled, the concurrency starts very high due to a large number of requests requiring only short responses. Figure \ref{fig:running_ab} (c) compares the impact of enabling prefix caching. With prefix caching, the concurrency quickly reaches a high level, whereas without prefix caching, the concurrency increases more gradually.

\begin{figure}[ht]
  \centering
  \begin{subfigure}[t]{0.74\linewidth}
    \centering
    \includegraphics[width=\linewidth, ]{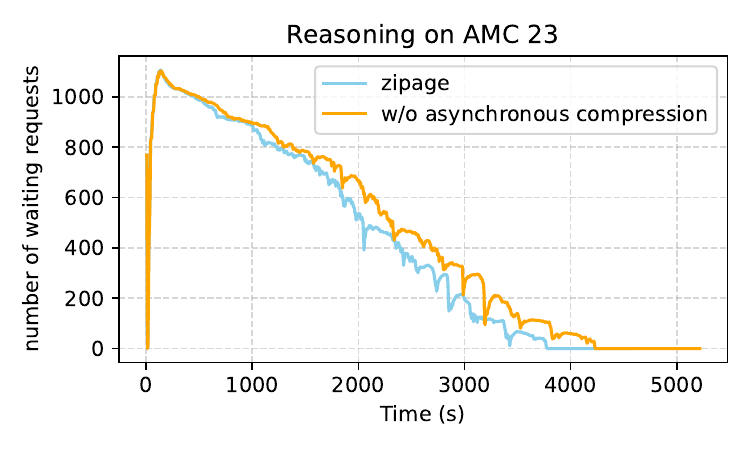}
    \caption{}
  \end{subfigure}

  \begin{subfigure}[t]{0.74\linewidth}
    \centering
    \includegraphics[width=\linewidth]{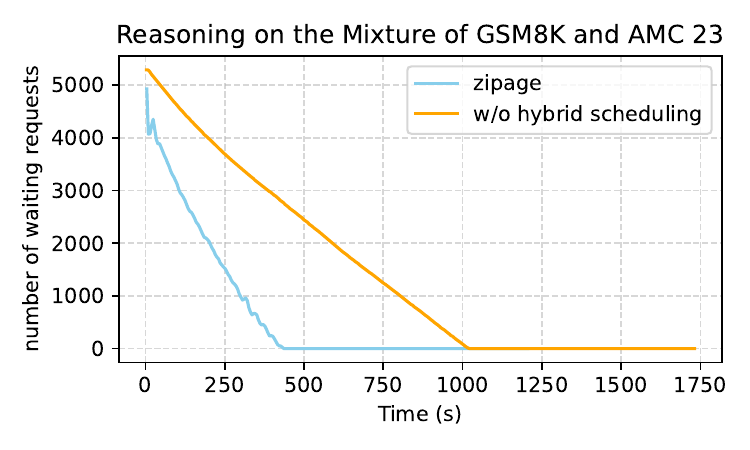}
    \caption{}
  \end{subfigure}
  \begin{subfigure}[t]{0.74\linewidth}
    \centering
    \includegraphics[width=\linewidth]{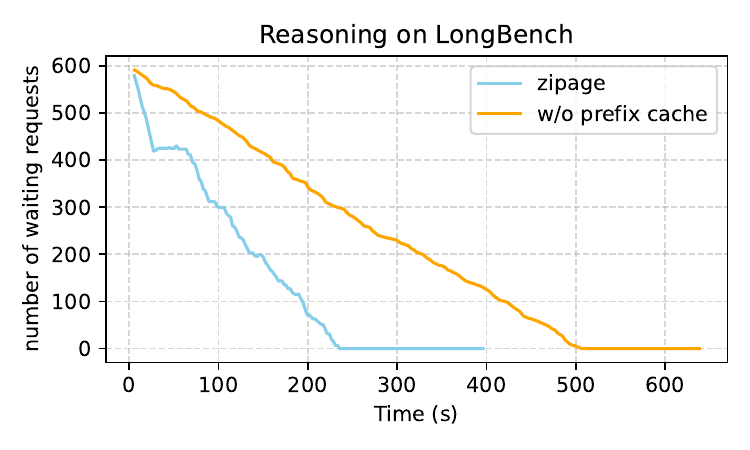}
    \caption{}
  \end{subfigure}

  \caption{Number of waiting requests during inference.}
  \label{fig:waiting_ab}
\end{figure}

Figure \ref{fig:waiting_ab} illustrates the number of waiting requests during inference. When the waiting queue is non-empty, the inference engine operates at full scheduling capacity. In this case, the slope of the waiting request curve indicates the request processing speed. Asynchronous compression, hybrid scheduling, and prefix caching all provide significant acceleration.

\begin{figure}[ht]
  \centering
  \begin{subfigure}[t]{0.74\linewidth}
    \centering
    \includegraphics[width=\linewidth, ]{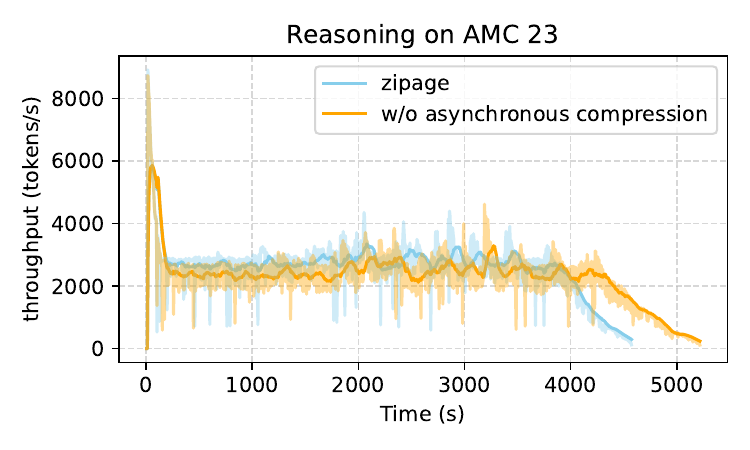}
    \caption{}
  \end{subfigure}

  \begin{subfigure}[t]{0.74\linewidth}
    \centering
    \includegraphics[width=\linewidth]{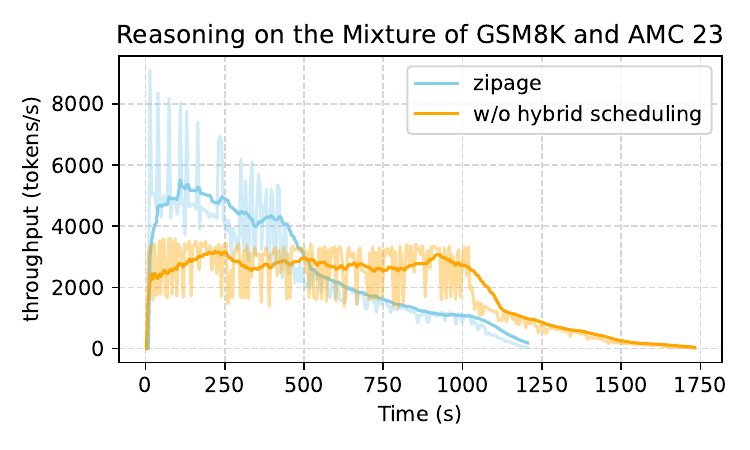}
    \caption{}
  \end{subfigure}
  \begin{subfigure}[t]{0.74\linewidth}
    \centering
    \includegraphics[width=\linewidth]{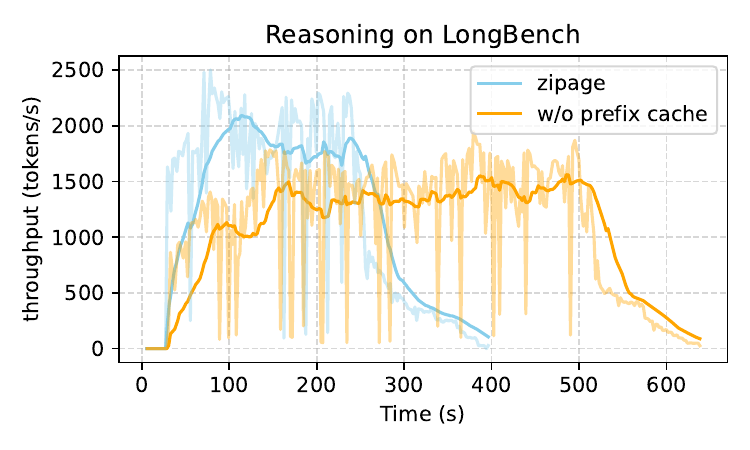}
    \caption{}
  \end{subfigure}
  \caption{Throughput during inference.}
  \label{fig:troughput_ab}
\end{figure}

Figure \ref{fig:troughput_ab} illustrates the real-time throughput during inference. Under request saturation, asynchronous compression, hybrid scheduling, and prefix caching deliver significant throughput improvements across various scenarios.

\begin{figure}[ht]
  \centering
  \begin{subfigure}[t]{0.74\linewidth}
    \centering
    \includegraphics[width=\linewidth]{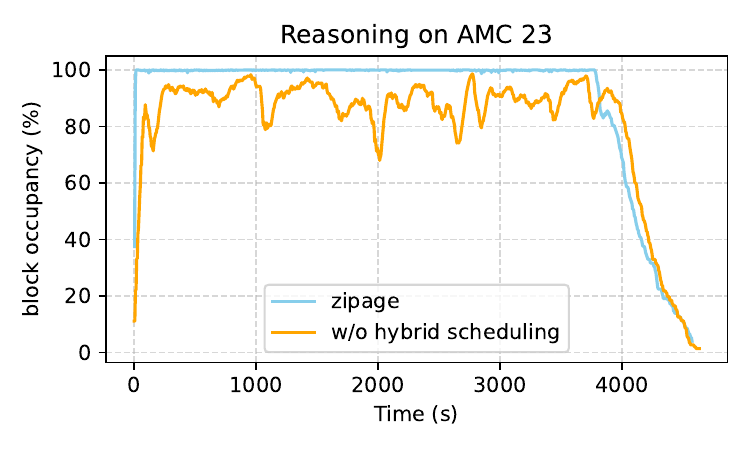}
    \caption{}
  \end{subfigure}
  \begin{subfigure}[t]{0.74\linewidth}
    \centering
    \includegraphics[width=\linewidth]{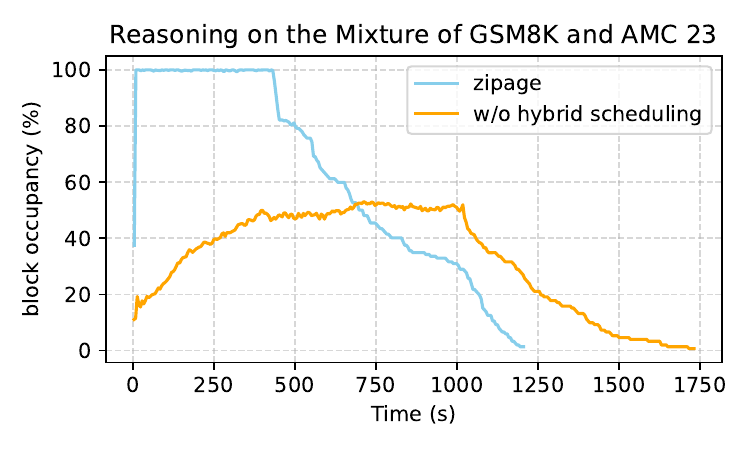}
    \caption{}
  \end{subfigure}
    \caption{Block occupancy during inference.}
  \label{fig:occupancy_ab}
\end{figure}

As previously mentioned, constrained scheduling limits concurrency, resulting in some blocks being underutilized. Figure \ref{fig:occupancy_ab} further illustrates the block utilization rate. On the mixed workload, which includes many requests with both short inputs and outputs, this underutilization becomes more pronounced, with less than half of the blocks being utilized most of the time.

\section{Additional Information of Comparison with PagedAttention}

\label{sct:cp_info}
\begin{figure}[ht]
  \centering

  \begin{subfigure}[t]{0.74\linewidth}
    \centering
    \includegraphics[width=\linewidth]{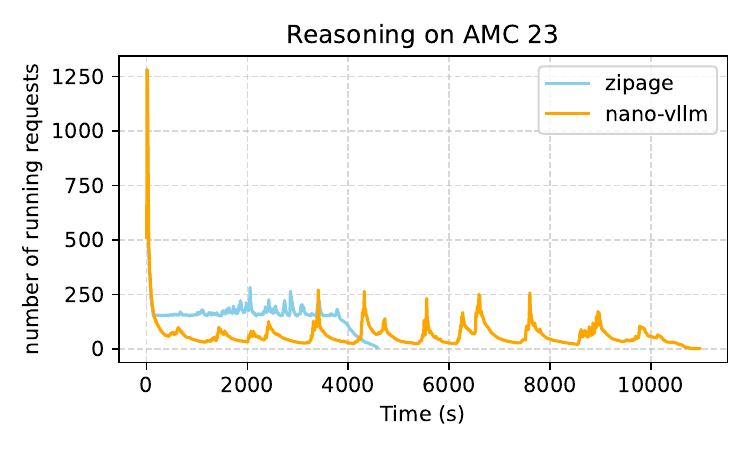}
    \caption{}
  \end{subfigure}
  \begin{subfigure}[t]{0.74\linewidth}
    \centering
    \includegraphics[width=\linewidth]{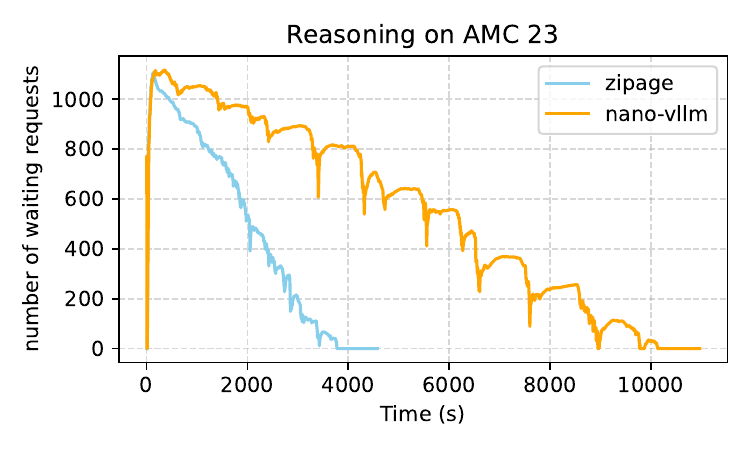}
    \caption{}
  \end{subfigure}
  \caption{The number of running and waiting requests during inference during inference on AMC23.}
  \label{fig:amc_cp}
\end{figure}

In this section, we provide additional comparative information between Zipage and Nano-vLLM. 

First, Figure \ref{fig:amc_cp} illustrates the number of running and waiting requests during inference on AMC23 using the Qwen3 model with different inference engines. The figure shows that the number of running requests in Nano-vLLM exhibits periodic fluctuations, consistent with the throughput variations in Figure \ref{fig:real_time} (a), while the step decoding time Figure in \ref{fig:real_time} (b) remains nearly constant. This indicates that the primary factor limiting throughput is concurrency, or more specifically, constrained concurrency caused by limited memory. In contrast to Nano-vLLM, Zipage maintains consistently high throughput throughout because, under the Zipage framework, the maximum memory usage per request is capped at a predefined limit, rather than continuously growing as the sequence length increases.

\begin{figure}[htp]
  \centering
  \begin{subfigure}[t]{0.74\linewidth}
    \centering
    \includegraphics[width=\linewidth]{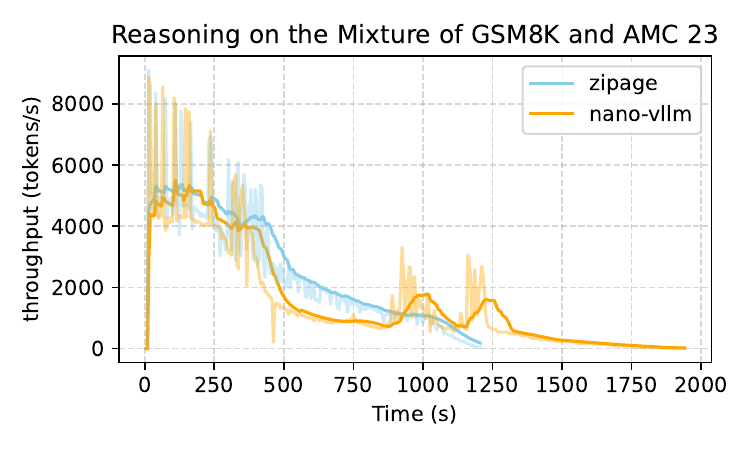}
    \caption{}
  \end{subfigure}

  \begin{subfigure}[t]{0.74\linewidth}
    \centering
    \includegraphics[width=\linewidth]{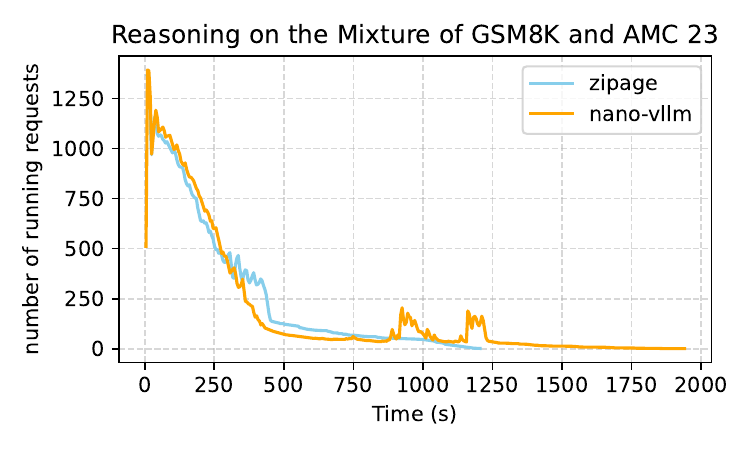}
    \caption{}
  \end{subfigure}
  \begin{subfigure}[t]{0.74\linewidth}
    \centering
    \includegraphics[width=\linewidth]{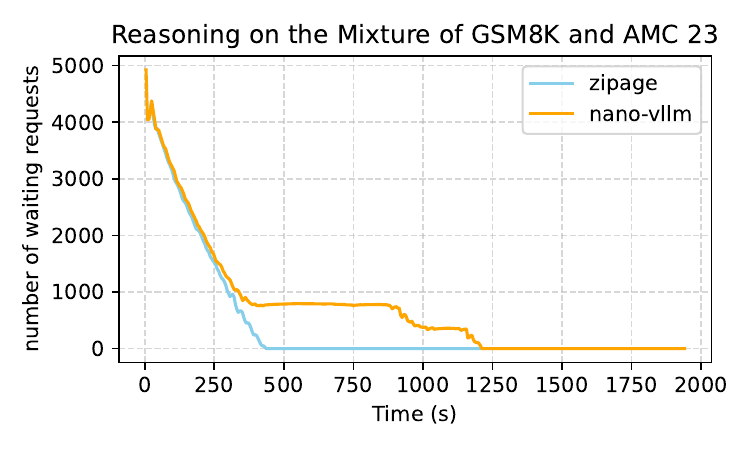}
    \caption{}
  \end{subfigure}
  \caption{The real-time throughput, number of running and waiting requests during inference during inference on the mixture of GSM8K and AMC 23.}
  \label{fig:mix_cp}
\end{figure}

Figure \ref{fig:mix_cp} illustrates the number of running and waiting requests during inference on the mixture of GSM8K and AMC 23 using the Qwen3 model with different inference engines. In this mixed workload, the initial performance of Zipage and Nano-vLLM is very similar, as short-response requests dominate at the beginning. However, as long-response requests occupy a large number of KV cache blocks, requests in Nano-vLLM's waiting queue experience longer delays, significantly increasing total execution time and reducing TPS. 

\begin{figure}[ht]
  \centering
  \begin{subfigure}[t]{0.74\linewidth}
    \centering
    \includegraphics[width=\linewidth]{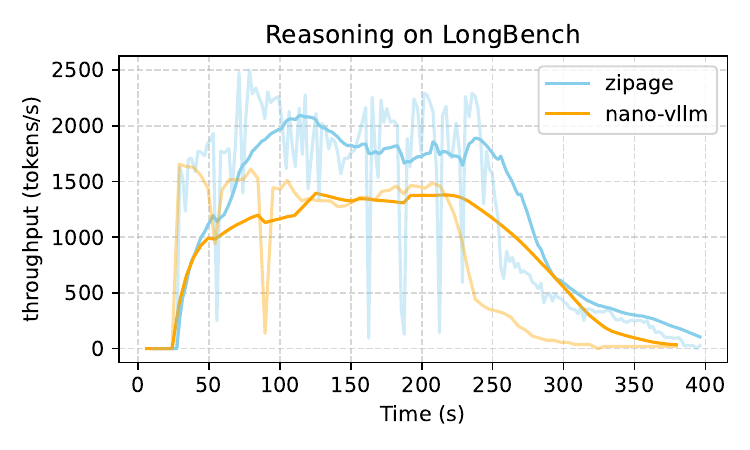}
    \caption{}
  \end{subfigure}

  \begin{subfigure}[t]{0.74\linewidth}
    \centering
    \includegraphics[width=\linewidth]{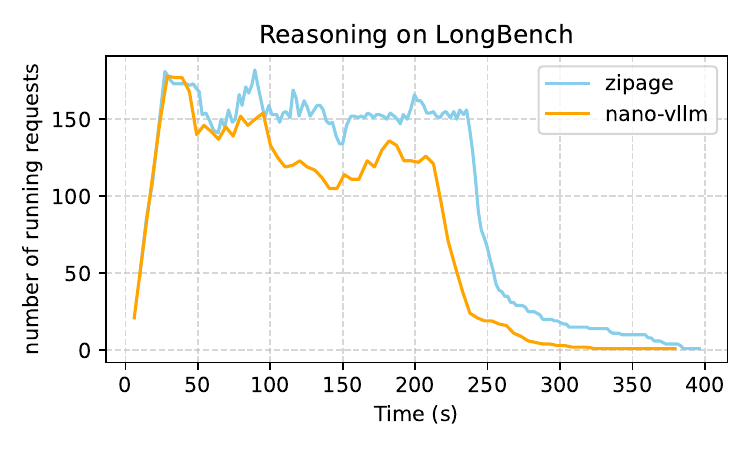}
    \caption{}
  \end{subfigure}
  \begin{subfigure}[t]{0.74\linewidth}
    \centering
    \includegraphics[width=\linewidth]{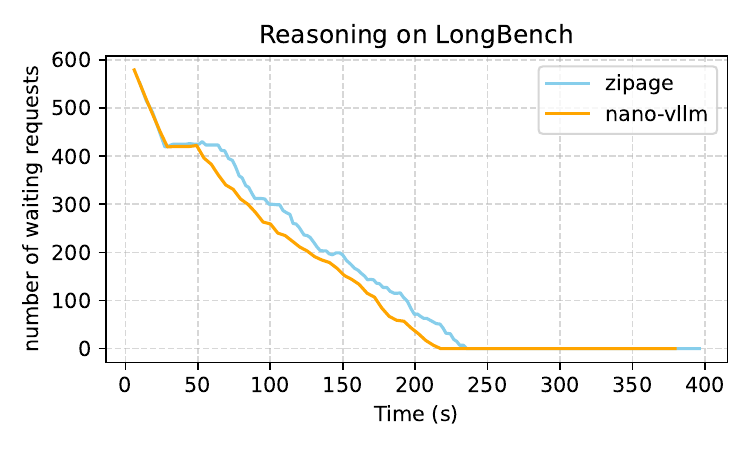}
    \caption{}
  \end{subfigure}
  \caption{The real-time throughput, number of running and waiting requests during inference during inference on LongBench.}
  \label{fig:long_cp}
\end{figure}

Figure \ref{fig:long_cp} illustrates the number of running and waiting requests during inference on LongBench using the Qwen3 model with different inference engines. We observe that both throughput and the number of running requests are higher with Zipage. However, the total time taken by Zipage to complete inference is slightly longer, likely due to some excessively long outputs. While the average output length on LongBench is only 400 tokens, occasional requests may produce significantly longer outputs, which can substantially impact the total time.

\section{Experiments on Models of Different Scales.}
\label{sct:scale_info}

\begin{figure}[htp]
    \centering
    \includegraphics[width=0.85\linewidth]{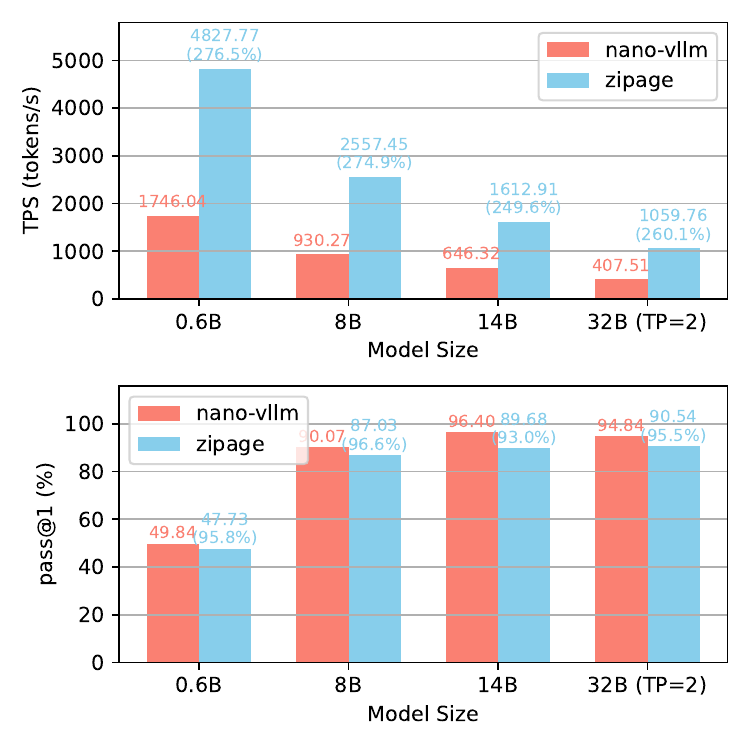}
    \caption{TPS and pass@1 performance across different model sizes. TP=2 indicates that the model is run on two GPUs using tensor parallelism, while all other experiments are conducted on a single GPU by default.}
    \label{fig:scale}
\end{figure}

In this section, we report experimental results for Qwen3 models of different sizes. For all experiments, Zipage uses a budget of 2048. Figure \ref{fig:scale} presents the TPS and pass@1 metrics for inference on AMC23 under the Zipage and Nano-vLLM frameworks. Across all model sizes, Zipage achieves significant throughput improvements. Additionally, the performance exceeds 95\% of Full KV for all sizes except 14B, which is slightly below 95\%.

\begin{figure}[ht]
  \centering

  \begin{subfigure}[t]{0.74\linewidth}
    \centering
    \includegraphics[width=\linewidth]{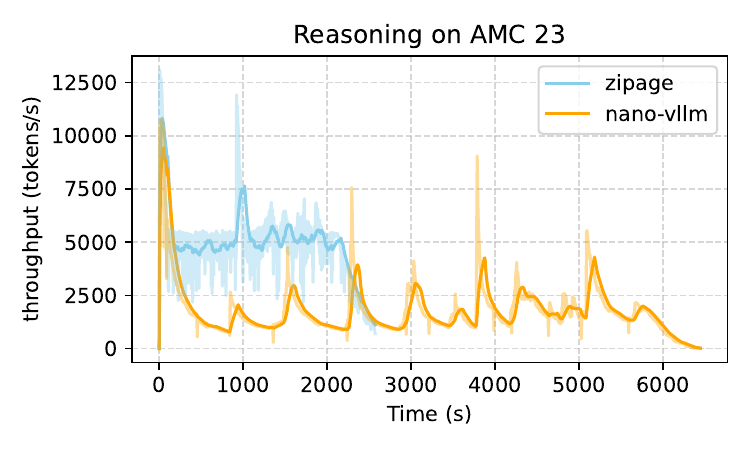}
    \caption{}
  \end{subfigure}
  \begin{subfigure}[t]{0.74\linewidth}
    \centering
    \includegraphics[width=\linewidth]{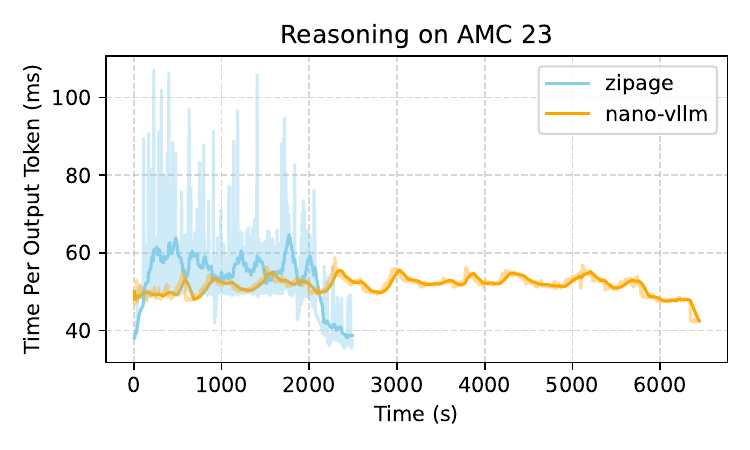}
    \caption{}
  \end{subfigure}

  \begin{subfigure}[t]{0.74\linewidth}
    \centering
    \includegraphics[width=\linewidth, trim=0 0 0 0, clip]{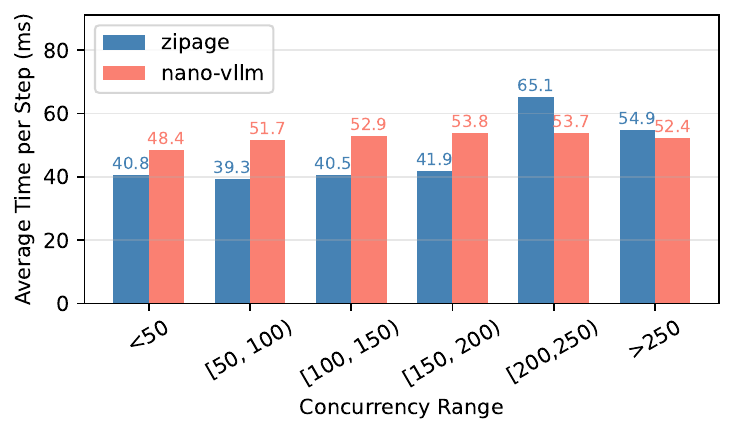}
    \caption{}
  \end{subfigure}
  \begin{subfigure}[t]{0.74\linewidth}
    \centering
    \includegraphics[width=\linewidth, trim=0 0 0 0, clip]{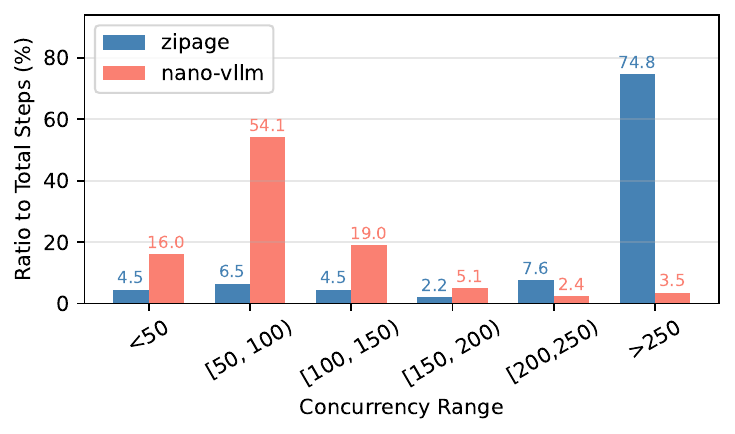}
    \caption{}
  \end{subfigure}

  \caption{The figure shows Qwen3 0.6B's performance using Zipage or Nano-vLLM on AMC 23, including:(a) real-time throughput, (b) per-step real-time decoding time, (c) average per-step time at different concurrency range, (d) and the ratio of steps to total steps under different concurrency range.}
  \label{fig:info_06b}
\end{figure}

\begin{figure}[ht]
  \centering
  \begin{subfigure}[t]{0.74\linewidth}
    \centering
    \includegraphics[width=\linewidth]{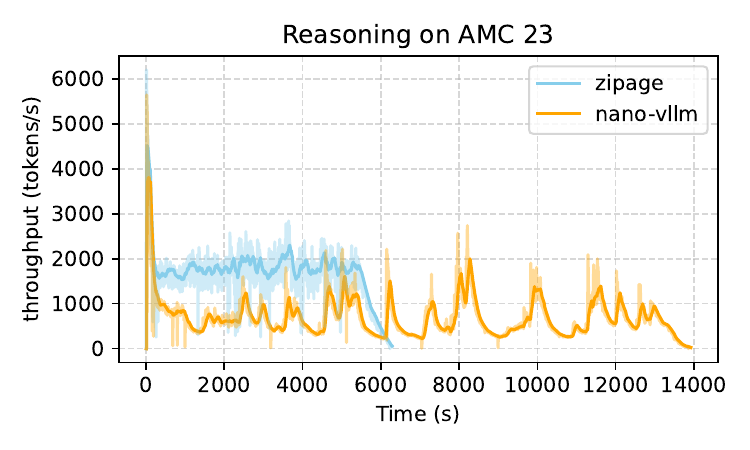}
    \caption{}
  \end{subfigure}
  \begin{subfigure}[t]{0.74\linewidth}
    \centering
    \includegraphics[width=\linewidth]{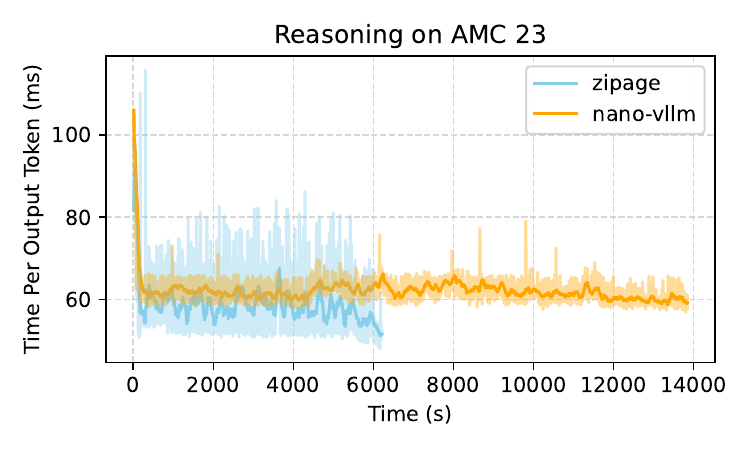}
    \caption{}
  \end{subfigure}
  \begin{subfigure}[t]{0.74\linewidth}
    \centering
    \includegraphics[width=\linewidth, trim=0 0 0 0, clip]{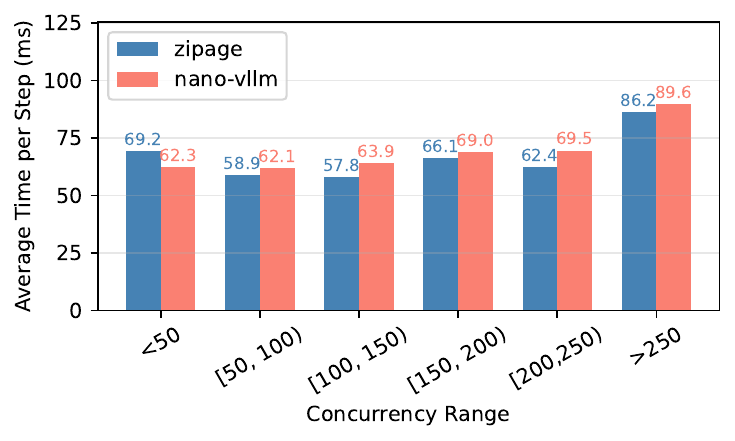}
    \caption{}
  \end{subfigure}
  \begin{subfigure}[t]{0.74\linewidth}
    \centering
    \includegraphics[width=\linewidth, trim=0 0 0 0, clip]{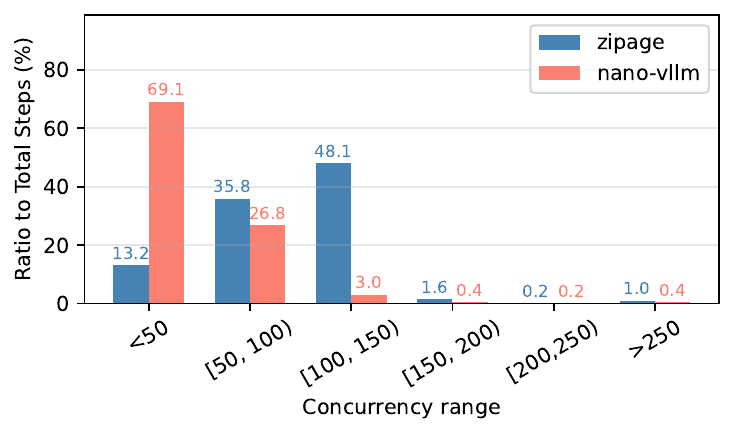}
    \caption{}
  \end{subfigure}
  \caption{The figure shows Qwen3 0.6B's performance using Zipage or Nano-vLLM on AMC 23, including:(a) real-time throughput, (b) per-step real-time decoding time, (c) average per-step time at different concurrency range, (d) and the ratio of steps to total steps under different concurrency range.}
  \label{fig:info_14b}
\end{figure}

\begin{figure}[ht]
  \centering
  \begin{subfigure}[t]{0.74\linewidth}
    \centering
    \includegraphics[width=\linewidth]{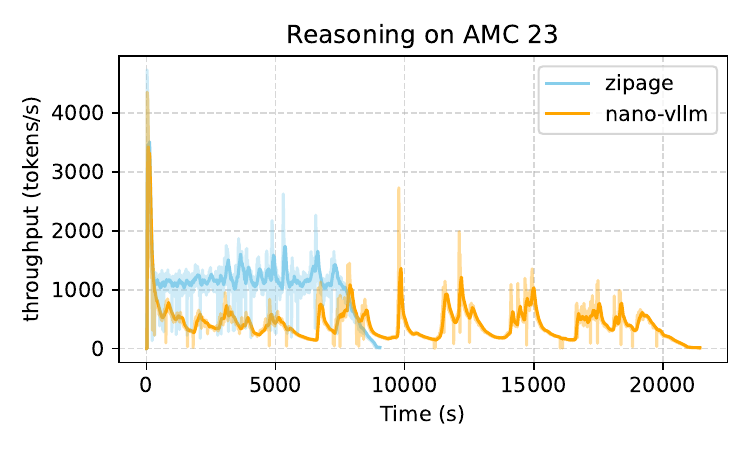}
    \caption{}
  \end{subfigure}
  \begin{subfigure}[t]{0.74\linewidth}
    \centering
    \includegraphics[width=\linewidth]{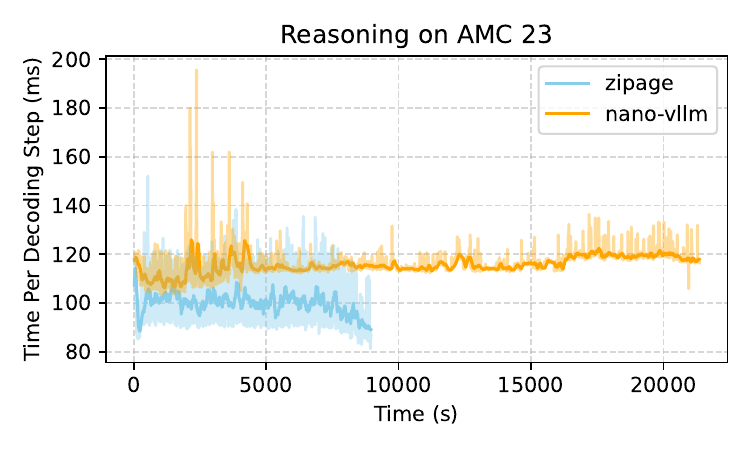}
    \caption{}
  \end{subfigure}
  \begin{subfigure}[t]{0.74\linewidth}
    \centering
    \includegraphics[width=\linewidth, trim=0 0 0 0, clip]{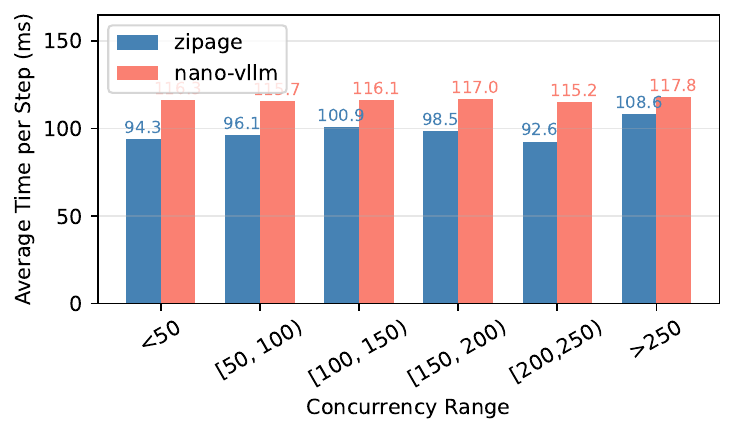}
    \caption{}
  \end{subfigure}
  \begin{subfigure}[t]{0.74\linewidth}
    \centering
    \includegraphics[width=\linewidth, trim=0 0 0 0, clip]{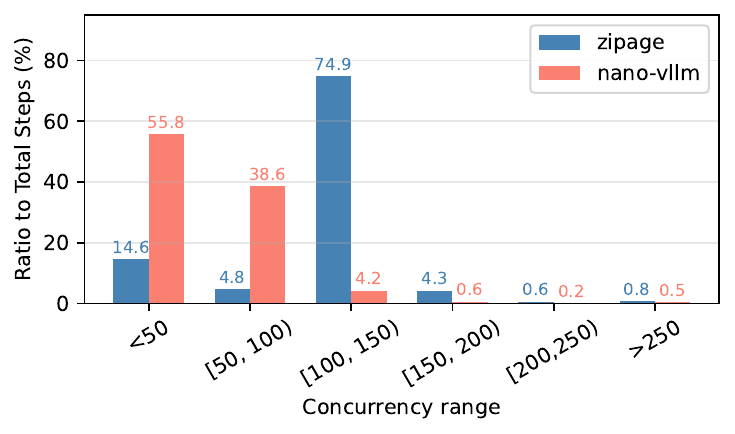}
    \caption{}
  \end{subfigure}
  \caption{The figure shows Qwen3 32B's  performance using Zipage or Nano-vLLM on AMC 23, including:(a) real-time throughput, (b) per-step real-time decoding time, (c) average per-step time at different concurrency range, (d) and the ratio of steps to total steps under different concurrency range. (Tensor parallelism = 2)}
  \label{fig:info_32b}
\end{figure}

\begin{figure}[ht]
  \centering
  \begin{subfigure}[t]{0.74\linewidth}
    \centering
    \includegraphics[width=\linewidth]{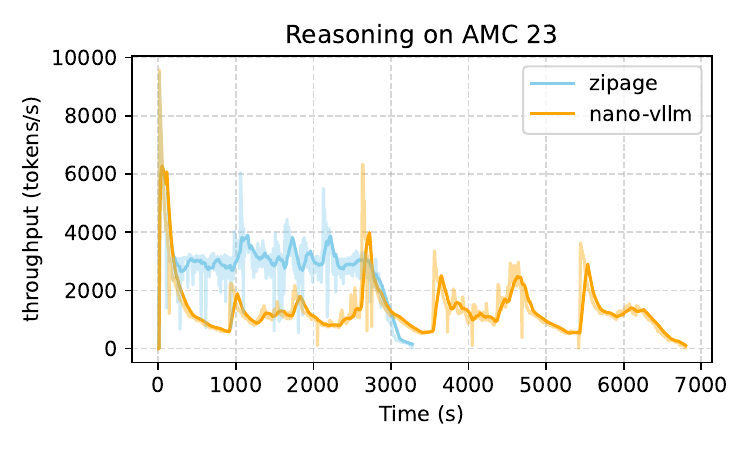}
    \caption{}
  \end{subfigure}
  \begin{subfigure}[t]{0.74\linewidth}
    \centering
    \includegraphics[width=\linewidth]{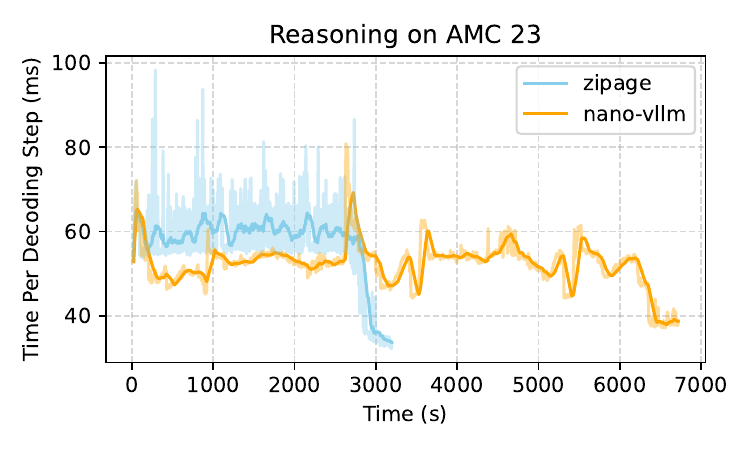}
    \caption{}
  \end{subfigure}

  \begin{subfigure}[t]{0.74\linewidth}
    \centering
    \includegraphics[width=\linewidth, trim=0 0 0 0, clip]{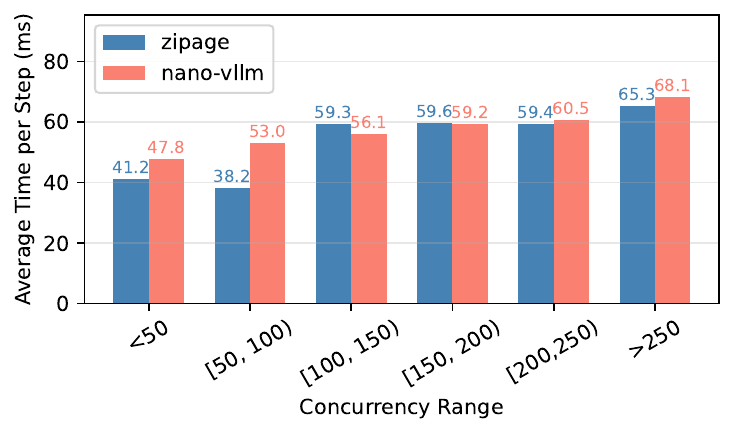}
    \caption{}
  \end{subfigure}
  \begin{subfigure}[t]{0.74\linewidth}
    \centering
    \includegraphics[width=\linewidth, trim=0 0 0 0, clip]{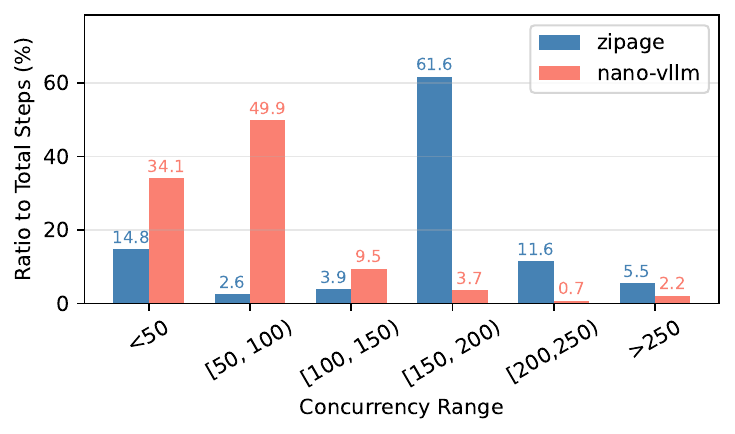}
    \caption{}
  \end{subfigure}
  \caption{The figure shows DS Llama 8B's  performance using Zipage or Nano-vLLM on AMC 23, including:(a) real-time throughput, (b) per-step real-time decoding time, (c) average per-step time at different concurrency range, (d) and the ratio of steps to total steps under different concurrency range.}
  \label{fig:info_llama8b}
\end{figure}

Real-time throughput, average decoding time per step, and the proportion across different concurrency ranges for the 0.6B, 14B, and 32B models are shown in Figures \ref{fig:info_06b}, \ref{fig:info_14b}, and \ref{fig:info_32b}, respectively. Figure \ref{fig:info_llama8b} provides details for the DS Llama 8B model.

\end{document}